\documentclass[12pt]{iopart}
\usepackage{amsfonts,amssymb,graphicx}
\usepackage{iopams} 
\usepackage{color}
\usepackage{float}

\def\be{\begin{equation}}
\def\ee{\end{equation}}
\def\bea{\begin{eqnarray}}
\def\eea{\end{eqnarray}}
\def\eq#1{(\ref{#1})}

\def\ms{\medskip}
\def\fig#1{figure \ref{#1}}
\def\figs#1{figures \ref{#1}}

\def\sec#1{section \ref{#1}}
\def\secs#1{sections \ref{#1}}

\def\bfr{{\bf r}}
\def\bfp{{\bf p}}

\def\d{{\rm d}}
\def\rb{r_{\rm b}}

\def\simg{\,\hbox{\kern.1em \lower.6ex \hbox{$\sim$} \kern-1.12em
          \raise.6ex \hbox{$>$} }}
\def\siml{\,\hbox{\kern.1em \lower.6ex \hbox{$\sim$} \kern-1.12em
          \raise.6ex \hbox{$<$} }}
\def\lambdab{\widetilde{\lambda}}

\newcommand{\ggt}{\widetilde{g}}
\begin{document}

\title[Closed orbits and spatial density oscillations in the circular billiard]
      {Closed orbits and spatial density oscillations\\ in the circular billiard}

\author[M Brack and J Roccia]{Matthias Brack and J\'er\^ome Roccia}

\address{Institute for Theoretical Physics, University of
Regensburg, D-93040 Regensburg, Germany}

\date{\today}

\begin{abstract} 
We present a case study for the semiclassical calculation of the
oscillations in the particle and kinetic-energy densities for the
two-dimensional circular billiard. For this system, we can give a complete 
classification of all closed periodic and non-periodic orbits. We discuss 
their bifurcations under variation of the starting point $r$ and derive
analytical expressions for their properties such as actions, stability determinants, 
momentum mismatches and Morse indices. We present semiclassical calculations of 
the spatial density oscillations using a recently developed closed-orbit theory 
[Roccia J and Brack M 2008 {\it Phys.\ Rev.\ Lett.} {\bf 100} 200408], employing 
standard uniform approximations from perturbation and bifurcation theory, 
and test the convergence of the closed-orbit sum.

\end{abstract}


%03.65.Sq 	Semiclassical theories and applications

%03.75.Ss 	Degenerate Fermi gases

%05.30.Fk 	Fermion systems and electron gas (see also 71.10.-w
%		 Theories and models of many-electron systems)

%71.10.-w 	Theories and models of many-electron systems

%71.10.Ca 	Electron gas, Fermi gas

%71.15.Mb 	Density functional theory, local density approximation,
%		gradient and other corrections

\maketitle

\section{Introduction}

We have recently developed \cite{rb,rbkm} a semiclassical theory
for spatial density os\-cillations in quantum systems consisting of 
$N$ non-interacting fermions bound in a local mean-field potential 
$V(\bfr)$. Our theory makes use of Gutzwiller's semi\-classical 
approximation of the single-particle Green function \cite{gutz} 
(which is essentially the Fourier transform of the semiclassical Van 
Vleck propagator \cite{vlec} but adding all extra phases resulting 
from integrations done in the stationary phase approximation).
While the average parts of the particle and
kinetic-energy densities are given by the extended Thomas-Fermi 
theory (see, e.g., \cite{book}, chapter 4), their oscillating 
parts could be expressed \cite{rb} as sums over all closed, in 
general non-periodic orbits of the corresponding classical system, 
starting and ending at the same space point $\bfr$. The semiclassical
evaluation of the spatial density oscillations requires the knowledge 
of all closed non-periodic orbits and their actions, stability
determinants, momentum mismatches and Morse indices (as specified in 
the next section), which in general can only be determined numerically.
The resulting formulae are the analogues of the semiclassical trace 
formulae for the level density, initiated by Gutzwiller \cite{gutz}, 
in terms of periodic orbits.

As an outcome of the semiclassical theory, some ``local virial
theorems'' connecting kinetic and potential energy densities at any 
given point $\bfr$ and some (integro-) differential equations for 
the particle density $\rho(\bfr)$, that previously had been proved
to hold exactly for isotropic harmonic oscillators with closed shells 
\cite{bm} and for linear potentials \cite{rbkm} in $D$ spatial 
dimensions, could be generalized to arbitrary local potentials. 
Formally, they hold in general only in the semiclassical limit 
$\hbar\to 0$, corresponding to $N\to\infty$, but in various model 
potentials they have been shown \cite{rbkm} to be well fulfilled 
even for moderate particle numbers $N$.

The present paper is a case study for the two-dimensional circular 
billiard, for which we can classify all closed orbits and determine 
their properties analytically, and that allows us to study the
convergence of the orbit sum in the semiclassical formulae for the
spatial density oscillations. In section \ref{secscl} we summarize
the main results of the semiclassical theory. In section \ref{secorb}
we investigate the closed orbits of the circular billiard. We give
a complete classification of all periodic and non-periodic orbits,
in\-cluding their bifurcations that can occur at specific distances
$\rb$ from the center, and present analytical expressions for their
properties required in the semiclassical theory. In section \ref{secregul}
we discuss the regularizations that become necessary at critical
points where the semiclassical amplitudes diverge for different
reasons. These are the center $r=0$ (symmetry breaking), the
bifurcation points $\rb$ ($0<\rb<R$), and the boundary $r=R$ 
(zero length of the shortest orbit). In section \ref{secnum} we
present results for the spatial density oscillations and test, 
in particular, their convergence and its relation to the shell 
effects in the total energy. In the appendix, we give explicit 
analytical results for the properties of some of the shortest orbits.

\section{Semiclassical closed-orbit theory}
\label{secscl}

In this section we present the general framework and the main 
results of the semiclassical theory for spatial density oscillations
developed in \cite{rb,rbkm}. 
We consider a $D$-dimensional system of $N$ non-interacting particles 
with mass $m$, obeying Fermi-Dirac statistics and bound by a local 
potential $V({\bf r})$. $V({\bf r})$ may re\-present the self-consistent 
local mean field of an {\it interacting} fermion system, such as it is
obtained in density functional theory, or just a given model potential. 
The discrete energy eigenvalues $E_j$ and eigenfunctions $\psi_j(\bfr)$ 
are given by the stationary Schr\"odinger equation. The
quantum-mechanical particle density of the system at zero temperature, 
including a factor two for the spin degeneracy, is given by
\begin{equation} 
\rho(\bfr)= 2\!\sum_{E_j \leq \lambda} \psi_j^{\star}({\bf r}) \psi_j(\bfr)\,,
\qquad \int \rho(\bfr)\,{\rm d}^Dr = N\,,
\label{rho}
\end{equation}
where the Fermi energy $\lambda=\lambda(N)$ is determined by the 
normalization of the density to the given particle number $N$ (which 
is here taken as an even integer). For the kinetic-energy density
we discuss two different forms
\begin{eqnarray}
\tau({\bf r}) & = & -\frac{\hbar^2}{2m}\;2\!\sum_{E_j \leq \lambda}
                    \psi_j^{\star}({\bf r}) \nabla^2\psi_j({\bf r}) \,,  
\label{tau}\\
\tau_1(\bfr)  & = & \frac{\hbar^2}{2m}\;2\!\sum_{E_j \leq \lambda}
                    |\nabla \psi_j({\bf r})|^2  \,, 
\label{tau1}
\end{eqnarray}
which upon integration both lead to the exact total kin\-etic energy.
We also investigate the average of these two kinetic-energy densities:
\begin{equation} 
\xi(\bfr)= \frac12\,[\tau(\bfr)+\tau_1(\bfr)]\,.
\label{xi}
\end{equation}
Due to the assumed time reversal symmetry and spin degeneracy, these
densities are interrelated by the expressions
\be\hspace{-.5cm}
\tau(\bfr) = \xi(\bfr)-\frac14\,\frac{\hbar^2}{2m}\,\nabla^2\rho(\bfr)\,,\qquad      
\tau_1(\bfr) = \xi(\bfr)+\frac14\,\frac{\hbar^2}{2m}\,\nabla^2\rho(\bfr)\,.
\label{tauxi}
\ee

We now separate all densities into a smooth and an oscillating part:
\bea
\rho(\bfr)&=&{\widetilde\rho}(\bfr)+\delta\rho(\bfr),\\
\tau(\bfr)&=&{\widetilde\tau}(\bfr)+\delta\tau(\bfr), \\
\tau_1(\bfr)&=&{\widetilde\tau}_1(\bfr)+\delta\tau_1(\bfr), \\
\xi(\bfr)&=&{\widetilde\xi}(\bfr)+\delta\xi(\bfr).
\label{densep}
\eea
The smooth parts are for differentiable potentials $V(\bfr)$ given by the extended 
Thomas-Fermi (ETF model) in terms of gradients of $V(\bfr)$.
For billiard systems, where no gradient expansion of the potential exists,
they are just given by their constant TF values. For the oscillating 
parts we have derived \cite{rb} a semiclassical expansion in terms 
of classical orbits which shall be examined analytically here for 
the circular billiard with radius $R$. 

The potential of the circular billiard is defined by
\be
V=0\, \quad \hbox{for} \quad 0 \leq r \leq R\,,\qquad
V=\infty \quad \hbox{for} \quad r>R\,,
\label{box}
\ee
where $(r,\phi)$ are polar coordinates. We solve the Schr\"odinger equation 
for this potential with the Dirichlet boundary condition $\psi_j(r=R,\phi)=0$. 
$\{j\}$ are given by the set of the radial quantum numbers $n=0,1,2,\dots$ and 
the angular-momentum quantum numbers $l=0,\pm 1,\pm2,\dots$ The eigenenergies 
and normalized wave functions are given \cite{rob} by
\bea\hspace{.9cm}
E_{nl}=\; z_{nl}^2 E_0\,,\qquad E_0=\hbar^2\!/(2 m R^2)\,,\nonumber\\
\psi_{nl}(r,\phi)=\; c_{nl} J_l\left(z_{nl}\,r/R\right)e^{ i l \phi}.
\label{eig}
\eea
The normalization constants are 
$c_{nl}=[\sqrt{\pi}R J_{l+1}(z_{nl})]^{-1}$, where $z_{nl}$ is the $n$-th zero 
of the cylindrical Bessel function $J_l(z)$. $\psi_{nl}$ can be summed to yield 
the exact quantum-mechanical densities \eq{rho} -- \eq{tau1}. 

The smooth parts of the densities, which in billiard systems are
independent of $r$, are given by their TF 
values\footnote{We note that the density ${\rho_{\rm TF}}$ in \eq{tfden}
cannot be normalized to the correct particle number $N$. The reason 
is that the constant TF density is not able to reproduce the sharp
decrease of the quantum-mechanical density $\rho(r)$ near the boundary
$r=R$, where it is forced to become zero.}
in terms of a smooth Fermi energy $\lambdab$:
\be
\hspace{-1.5cm}
{\tilde\rho} = \rho_{\rm TF} = \frac{m\lambdab}{\pi\hbar^2}
                             = \frac{1}{2\pi R^2}\left(\frac{\lambdab}{E_0}\right),\qquad
{\tilde\tau} = \tau_{\rm TF} = \frac{m\lambdab^2}{2\pi\hbar^2}
                             = \frac{1}{4\pi R^2}\left(\frac{\lambdab^2}{E_0}\right).
\label{tfden}
\ee
Note that $\tilde{\tau}$ is the common smooth part of all three kinetic 
energy densities \eq{tau} -- \eq{xi}. From the Weyl expansion \cite{bh,beho} 
of the integrated level density given in \eq{NofE}, we can find the 
asymptotic expansion of the Fermi energy $\lambdab(N)$ in powers of $N^{-1/2}$ :
\be
\lambdab(N) = E_0 \bigg[2N+2(2N)^{1/2}+\frac{4}{3}+ {\cal O}(N^{-1/2})\bigg]\,.
\label{lambofn}
\ee

For the oscillating parts of the densities, we now reproduce the main 
formulae from \cite{rb} for the special case of $D=2$ space dimensions 
and spherical symmetry when all densities only depend on the radial 
variable $r=|\bfr|$. To leading order in $\hbar$, the semiclassical 
expressions for the oscillating parts of the densities are given by
\begin{eqnarray}
\hspace{-1.cm}\delta \rho(r)   \simeq  \,\sum_\gamma {\cal A}_\gamma(\lambdab,r)
                           \,\cos\left[\frac{1}{\hbar}S_\gamma(\lambdab,r)
                           -\mu_\gamma\frac{\pi}{2}-\frac{3\pi}{4}\right],
\label{drhosc}\\
\hspace{-1.cm}\delta \tau(r)   \simeq
\,\frac{p_\lambda^2}{2m}\,\sum_\gamma 
                           {\cal A}_\gamma(\lambdab,r)
                           \,\cos\left[\frac{1}{\hbar}S_\gamma(\lambdab,r)
                           -\mu_\gamma\frac{\pi}{2}-\frac{3\pi}{4}\right],  
\label{dtausc}\\
\hspace{-1.cm}\delta\tau_1(r) \simeq  \,\frac{p_\lambda^2}{2m}\,\sum_\gamma 
                           {\cal A}_\gamma(\lambdab,r)\,Q_\gamma(\lambdab,r)
                           \,\cos\left[\frac{1}{\hbar}S_\gamma(\lambdab,r)
                           -\mu_\gamma\frac{\pi}{2}-\frac{3\pi}{4}\right].  
\label{dtau1sc}
\end{eqnarray}
The sum is over all closed orbits $\gamma$ starting and ending in the point $r$. 
For {\it periodic orbits} (POs) the action integral $S_\gamma$
becomes independent of $r$. Therefore, the POs do not yield any oscillating
phases in the above expressions and their contributions are smooth functions 
of $r$ varying only through the amplitude factors ${\cal A}_\gamma$ and 
$Q_\gamma$ in the expressions above. The leading contributions to the
density oscillations therefore come from the {\it non-periodic orbits} 
(NPOs). In one-dimensional systems it has, in fact, been shown \cite{rb,rbkm} 
that the contributions of the POs are completely absorbed by the
smooth TF parts of the densities. In higher-dimensional systems, 
like the present circular billiard, some contributions of POs
must be included in connection with symmetry breaking at $r=0$ and
with bifurcations at finite distances $r>0$, as will be discussed in
section \ref{secregul}. 

The action function $S_\gamma(\lambdab,r)=S_\gamma(\lambdab,r,r'=r)$ 
is gained from the general open action integral for an orbit starting
at $\bfr$ and ending at $\bfr'$ at fixed energy $E=\lambdab$
\be
S_\gamma(\lambdab,\bfr,\bfr') = \int_{\bfr}^{\bfr'} {\bf p}(\lambdab,{\bf q})
                                \cdot \d\,{\bf q}\,,
\label{actint}
\ee
where ${\bf p}(\lambdab,\bfr)$ is the classical momentum in the point $\bfr$,
for $V(\bfr)=0$ given by
\be
{\bf p}(\lambdab,\bfr) = \frac{\dot{{\bf r}}}{|{\dot{\bf r}}|}\sqrt{2m\lambdab}\,,
\label{pclass}
\ee
whose modulus is a constant of motion and denoted here by 
$p_\lambda=|{\bf p}(\lambdab,\bfr)|$. $\mu_\gamma$ is the Morse index 
that counts the number of conjugate points along the orbit \cite{gutz} and will 
be discussed in section \ref{secmors}. The quantity $Q_\gamma(\lambdab,r)$ 
appearing in \eq{dtau1sc} for $\delta\tau_1(r)$ is defined as
\be
Q_\gamma(\lambdab,r) = \frac{[{\bf p}(\lambdab,\bfr)
                       \cdot{\bf p}(\lambdab,\bfr')]_{\gamma,\,\bfr'=\bfr}}{p_\lambda^2} 
                     = \cos[\,\theta({\bfp,\bfp'})]_\gamma\,,
\label{mismatch}
\ee
where $\bfp$ and $\bfp'$ are the short notations for the initial and 
final momentum, respectively, of the orbit $\gamma$ at the point $\bfr$. 
These are obtained also from the action integral \eq{actint} by
\be
\left. \nabla_{\bfr}S_\gamma(\lambdab,{\bf r,r'})\right|_{\bfr=\bfr'} = -\bfp\,,\quad
\left. \nabla_{\bfr'}S_\gamma(\lambdab,{\bf r,r'})\right|_{\bfr=\bfr'} = \bfp'\,.
\label{pcanon}
\ee
Since $Q_\gamma(\lambdab,r)$ in \eq{mismatch} depends on the angle $\theta$ between $\bfp$ 
and $\bfp'$, it may be called the ``momentum mismatch function'', being $Q_\gamma=+1$ for 
POs with $\bfp=\bfp'$ and $Q_\gamma=-1$ for self-retracing NPOs with $\bfp=-\bfp'$. 

The common semiclassical amplitude ${\cal A}_\gamma(\lambdab,r)$ in all densities is given by
\be
{\cal A}_\gamma(\lambdab,r) = \frac{2m}{\pi p_\lambda}\,
                              \frac{\sqrt{|{\cal D}_{\bot\gamma}(\lambdab,r)|}}
                              {\sqrt{2\pi\hbar}\;T_\gamma(\lambdab,r)}\,, \qquad \left.
{\cal D}_{\bot\gamma}(\lambdab,r) =
       \left(\frac{\partial{p}_{\bot}}{\partial{r'_{\!\bot}}}\right)\right|_{\bfr'=\bfr}.
\label{amp1}
\ee
Here $T_\gamma(\lambdab,r)={\rm d}S_\gamma(E,r)/{\rm d}E|_{E=\lambdab}$ is the 
running time of the orbit, and ${\cal D}_{\bot\gamma}$ its stability determinant
calculated from the components $p_{\bot}$ and $r'_{\!\bot}$ {\it transverse} 
to the orbit $\gamma$ of the initial momentum and final coordinate, respectively.
We shall in the following express the amplitudes defined in \eq{amp1} through
the Jacobians ${\cal J}_\gamma$, which are defined as the inverse
stability determinants, and omit the Fermi energy $\lambdab$ from all arguments:
\be
{\cal A}_\gamma(r) = \frac{2m}{p_\lambda\pi\sqrt{2\pi\hbar}\;T_\gamma(r)}
                     \frac{1}{\sqrt{|{\cal J}_\gamma(r)|}}\,, \qquad \left.
{\cal J}_\gamma(r) = \left(\frac{\delta r'_\perp}{\delta p_\perp}\right)\right|_{\bfr'=\bfr}\!.
\label{amp}
\ee

From \eq{drhosc} and \eq{dtausc} using $p_\lambda^2/2m=\lambdab$, we find
immediately the semiclassical relation
\be
\delta\tau(r) \simeq \lambdab\,\delta\rho(r)\,,
\label{lvt}
\ee
which is the {\it local virial theorem} discussed extensively in \cite{rbkm}
(valid here for $V(\bfr)=0$). Note, however, that this theorem fails
at the boundary $r=R$ (cf.\ \sec{secfried}).

It has been observed in \cite{rbkm} (see also \figs{figdisk4917} and
\ref{2ddisk4917}) that for systems with spherical symmetry for
$D>1$ (except harmonic oscillators), there exist two types
of oscillations:\\ ($i$) regular, short-ranged oscillations due to the
orbits librating in the radial direction, which we will briefly call
the {\it radial orbits}, and\\ ($ii$) irregular, long-ranged
oscillations due to the {\it non-radial orbits}.
(Note that in isotropic harmonic oscillators, all non-radial orbits 
are periodic, which explains the fact \cite{rbkm,bm} that these 
systems do not exhibit any irregular oscillations.) 
Like in \cite{rbkm}, we therefore decompose the oscillating 
parts of the densities \eq{drhosc} -- \eq{dtau1sc} in the following way 
(in obvious notation)
\bea
\delta\rho(r) & = & \delta_{\rm r}\rho(r) + \delta_{\rm irr}\rho(r)\,,\nonumber\\
\delta\tau(r) & = & \delta_{\rm r}\tau(r) + \delta_{\rm irr}\tau(r)\,,\nonumber\\
\delta\tau_1(r) & = & \delta_{\rm r}\tau_1(r) + \delta_{\rm irr}\tau_1(r)\,.
\label{seden}
\eea

The contribution of the radial NPOs to the particle density
\eq{drhosc} near $r=0$ can be summed over all repetitions $k$ and has 
already been given in \cite{rb,rbkm}. For the present system it becomes:
\begin{equation}
\delta_{\rm r}\rho(r) = (-1)^{^{M\!-1}}\!\frac{m}{\hbar\,T_{\rm r 1}(\lambdab)}
                    \,J_0(2rp_\lambda/\hbar)\,.   \qquad (r\ll R)
\label{delrhorad}
\end{equation}
Here $J_0(z)$ is the cylindrical Bessel function of order zero, $M$ is the 
number of filled ``main shells'' to be discussed in \sec{secnum}, and 
$T_{\rm r1}=4mR/p_\lambda$ is the period of the diametrical PO. This result
will be generalized to larger radii in \sec{secsymbr}. 
The regularity of the short-ranged oscillations is due to the fact
that all radial NPOs including their repetitions $k$ (see their
discussion in \sec{seclin}) contribute with the same period in $r$
at all distances except near the boundary $r=R$. This is different 
for the non-radial NPOs (which will be discussed in \sec{secnlin}): 
their periods in $r$ depend on their individual forms and become
larger in the limit $r\to 0$ (cf.\ also \sec{secregnlin}), which
explains the irregular nature of the long-ranged oscillations.

The contribution of the radial NPOs to the kinetic energy
densities are then easily obtained from the general equations
\eq{drhosc} -- \eq{dtau1sc}. For $\delta_{\rm r}\tau(r)$ we may just
use the local virial theorem \eq{lvt} to get an expression valid
for small $r$. Since $Q_\gamma(r)=-1$ for all radial NPOs, we get the 
relation \cite{rb}
\be
\delta_{\rm r}\tau_1(r) = -\delta_{\rm r}\tau(r)\,,  \qquad (r\ll R)
\label{tautau1}
\ee
again valid for small $r$.

In the following section, we shall classify all closed orbits in the 
circular billiard and derive their analytical properties introduced above.

\newpage

\section{Closed orbits in the circular billiard}
\label{secorb}

\subsection{Classification and symmetries of orbits}
\label{secclass}

We may classify the NPOs in the circular billiard in the same way as it 
has been done for the POs by Balian and Bloch \cite{babl}. 
They specified them by pairs of integers $(v,w)$, where $v$ is the 
number of reflections at the boundary and $w$ is the winding number 
around the center ($r=0$). If $v$ and $w$ have a common divisor $k>1$, 
the number $k$ is the number of repetitions of the primitive orbit 
(which by definition has $k=1$). The librating diametrical POs 
have by definition the winding number $v=w$. For the non-radial NPOs, 
there generally exist no repetitions (i.e., $k=1$), except at the 
points $r=0$ and $r=R$ where they become periodic. Repetitions of 
non-radial NPOs do occur at specific isolated points $0<r<R$ at 
which they become identical with POs or fractions thereof. In the 
calculation of the semiclassical densities we found, however, that these 
repetitions are practically negligible because they lead to rather
long orbits. For the radial NPOs, a generalized repetition 
number $k\geq 0$ will be defined in \sec{secnlin}.

All orbits starting and ending in a fixed point $r>0$ are isolated and
occur in discrete degenerate pairs corresponding to the time-reversal 
symmetry. This holds also for the librating (i.e., radial)
POs.\footnote{This is different in the trace formula for the level 
density: there one integrates over all $r$ and thereby includes 
automatically all pairs of librations starting in opposite directions, 
so that librations must only be counted once}
All orbits starting and ending at the origin $r=0$ form families of
continuously degenerate orbits, since they can be rotated about arbitrary
finite angles around the origin due to the U(1) symmetry of the system. This
has consequences for their semiclassical amplitudes, as discussed in \sec{secsymbr}.

\subsection{Geometry of an arbitrary orbit with $v$ reflections and winding number $w$}
\label{secgeom}

\begin{figure}[t]
\begin{center}
\begin{minipage}{1.\linewidth}
\hspace{2.7cm}
\includegraphics[width=0.6\columnwidth,clip=true]{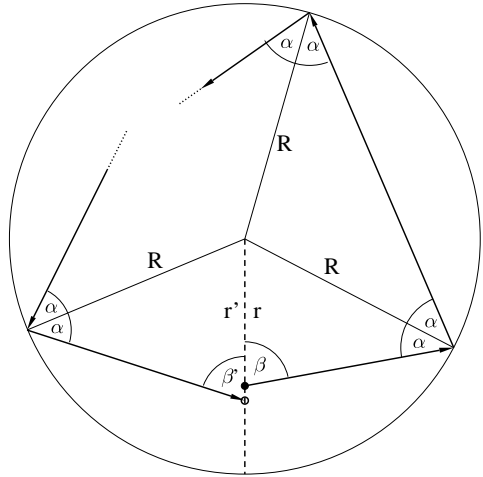} \vspace*{-0.5cm}
\end{minipage}
\end{center}
\caption{\label{circle}
Geometry of an arbitrary orbit in the circular billiard with radius $R$, 
starting at distance $r$ (dot) and ending at distance $r'$ (circle) from 
the center along the same radius vector (shown by the vertical dashed line).
}
\end{figure}

We now study the geometry of an arbitrary orbit, as illustrated in
\fig{circle}. Let the particle start from a point $r$ (black dot)
on a radius vector $0<r<R$ (vertical dashed line). Let $\beta$ be the 
starting angle with respect to that radius vector. Let $\beta'$ be the 
final angle at which the particle returns to the point $r'$ (circle)
on the same radius vector after $v$ reflections at the boundary and 
winding $w$ times around the origin. Both $\beta$ and $\beta'$ are 
chosen to lie {\it inside} the polygon formed by the orbit. The 
relation between $\beta$, $\beta'$ and the reflection angle $\alpha$ 
at the boundary then is
\be
\beta' = (1-w)\,2\pi + (v-1)\,\pi - 2v\alpha - \beta\,.
\label{betal}
\ee
Note that according to our definition, all angles $\alpha$, $\beta$ and $\beta'$ 
are non-negative. Moreover, they are restricted in all cases by $0\leq\alpha 
< \pi/2$ and $0\leq\beta,\beta'\leq\pi$. The lengths $r$ and $r'$ can be expressed
in terms of these angles and $R$ by
\be
r  = R\,\frac{\sin\alpha}{\sin\beta}\,,
\label{rofang}
\ee
\be
r' = R\,\frac{\sin\alpha}{\sin\beta'}\,.
\label{ropfang}
\ee

The condition for an orbit to be closed, $r=r'$, leads to $\sin\beta=\sin\beta'$ 
which has only two solutions in the physically interesting domain 
$\beta,\beta'\in[0,\pi]$, namely:\\
a) $\beta'=\beta$, giving in general a NPO (which, however, can 
become periodic in an isolated point where $\beta'=\beta=\pi/2$), and\\
b) $\beta'=\pi-\beta$, always giving a PO.

From \eq{rofang} we can express $\beta$ implicitly in terms of $r$ and
$\alpha$ by the following relations
\be\hspace{-1.cm}
\cos\beta=\pm\frac{R}{r}\,\sqrt{\frac{r^2}{R^2}-\sin^2\alpha}\,,\qquad
\cot\beta=\pm\frac{1}{\sin\alpha}\sqrt{\frac{r^2}{R^2}-\sin^2\alpha}\,.
\label{betofal}
\ee
The signs above must be taken to be positive if $\beta\leq\pi/2$ and 
negative if $\beta\geq\pi/2$.

\subsection{Non-periodic orbits}
\label{secNPOs}

Inserting $\beta'=\beta$ into \eq{betal}, the relation between $\beta$ 
and $\alpha$ becomes
\be
\beta = (1-w)\pi + (v-1)\pi/2 - v\alpha\,.
\label{betanpo}
\ee
Inserting \eq{betanpo} into \eq{rofang} yields $r$ as a function of
$\alpha$:
\be
r(\alpha) = (-1)^wR\,\frac{\sin\alpha}{\cos(v\pi/2-v\alpha)}\,,
\label{rofal}
\ee
which in general cannot be inverted explicitly. In order to find 
all NPOs in the circular billiard, it is sufficient to look for 
all real solutions $\alpha(r)$ of the inverse of equation \eq{rofal} in 
the physical range $0 \leq \alpha \leq \pi/2$. Upon variation of 
$r\in[0,R]$ for a fixed pair ($v,w$), U(1) symmetry breaking at $r=0$ creates 
non-radial NPOs, while bifurcations of some NPOs occur at specific 
points $r_{\rm b}>0$, whereby new NPOs or POs are created. It turns 
out that the radial NPOs and the diametric PO are the prime 
generators of many other closed orbits in the circular billiard. 
The remaining NPOs are created in tangent bifurcations. The  
bifurcation scenarios are discussed in detail in \sec{secnlin}. 

The NPOs are symmetric about the radius vector containing the
point $r$. They are doubly degenerate due to the time reversal 
symmetry, which here is identical to the reflection symmetry, 
except for $v=2$, $w=1$ in the limiting case $r=0$ ($\alpha=0$, 
$\beta=\pi$/2), in which case the orbit becomes radial and is 
mapped onto itself by time reversal. 

The length of the orbit $(v,w)$ becomes
\be
L_{v,w}(r) = 2(vR\cos\alpha+r\cos\beta)\,.
\label{Lvw}
\ee
The explicit dependence of \eq{Lvw} on $v$ and $w$ is obtained when
inserting \eq{betanpo} for $\beta$. Note that, because of the implicit $r$ 
dependence of $\alpha$ and $\beta$, the formula \eq{Lvw} does not reveal
the full (non-linear) dependence on the radius $r$. 

At $r=R$, the NPOs become periodic with $v+1$ or $v-1$ reflections (see the 
systematics in \sec{secnlin}). In this case we have $\beta=\alpha$ or
$\beta=\pi-\alpha$, and $L_{v,w}(r=R)=2R\,(v\pm1)\cos\alpha$. 
In the notation of Balian and Bloch \cite{babl}, the half polar angle 
covered between two successive reflections of an orbit with $v$ reflections
is $\phi_{v,w}=w\pi/v$. The lengths then become $L_{v,w}(r=R)=
2R\,(v\pm1)\sin\phi_{v\pm1,w}$. 

It is important to realize that the function $L_{v,w}(\alpha)$ given in 
\eq{Lvw} is {\it stationary with respect to $\alpha$}, at a fixed value 
of $r$, {\it precisely for the {\rm NPO}s obeying the relation} \eq{rofang}. Indeed, with 
the extra condition \eq{betanpo} which implies $\partial\beta/\partial\alpha=-v$, 
we find the stationary condition to lead to
\be  \left.
\frac{{\rm d}L_{v,w}(\alpha)}{{\rm d}\alpha}\right|_{r} = 2v\,(r\sin\beta-R\sin\alpha)=0\,,
\ee
which is identical with \eq{rofang}. For later reference, we give here also the second derivative:
\be  \left.
\frac{{\rm d}^2L_{v,w}(\alpha)}{{\rm d}\alpha^2}\right|_{r} = 
-2vR\cos\alpha\left(1+v\,\frac{r}{R}\,\frac{\cos\beta}{\cos\alpha}\right)\!.
\label{Lpp}
\ee

For the calculation of the Jacobian ${\cal J}_\gamma={\cal J}_{v,w}$ 
defined in \eq{amp}, we make a small variation of the initial momentum 
by a small variation $\delta\beta$ of the starting angle. The
reflection angle then will be $\alpha'=\alpha+\delta\alpha$. 
From \eq{rofang} with fixed $r$, we obtain
\be
\delta\alpha = \frac{r}{R}\,\frac{\cos\beta}{\cos\alpha}\,\delta\beta\,.
\label{deladelb}
\ee
The particle will then return to the point $r+\delta r'$ with the angle
$\beta'+\delta\beta'$, where
\be
\delta\beta' = -\delta\beta-2v\delta\alpha\,.
\label{delbetp}
\ee
From \eq{ropfang} we obtain the increment $\delta r'$ as
\be
\delta r' =
\frac{R}{\sin\beta}\,(\cos\alpha\,\delta\alpha-\sin\alpha\,
                                  \cot\beta\,\delta\beta').
\label{delrp}
\ee
Noting that the change of the starting transverse momentum is $\delta
p_\perp=p_\lambda\delta\beta$ and the change of the final transverse distance 
is $\delta r'_\perp = \delta r'\sin\beta$, we obtain the Jacobian 
${\cal J}_{v,w}$ as
\be
{\cal J}_{v,w}(r) = \frac{R}{p_\lambda}\left[\sin\alpha\,\cot\beta
                    +\frac{r}{R}\cos\beta\,(1+2v\,\tan\alpha\,\cot\beta)\right]\!.
\label{jacofang}
\ee
Using \eq{rofang} we can rewrite it in the form
\be
{\cal J}_{v,w}(r)  = \frac{2r}{p_\lambda}\,\cos\beta\left[1+v\,\frac{r}{R}\,
                     \frac{\cos\beta}{\cos\alpha}\right]\!.
\label{Jvw}
\ee
Like for the lengths \eq{Lvw}, the explicit dependence of \eq{jacofang} 
and \eq{Jvw} on $v$ and $w$ comes through \eq{betanpo}, and their full 
$r$ dependence is not revealed. At $r=R$ we obtain the values
\be
{\cal J}_{v,w}(r=R) = \frac{2R}{p_\lambda}\,(v\pm 1)\sin\phi_{v\pm1,w}\,.
\label{jnpoR}
\ee
%Note that \eq{Jvw} contains the same factor in the square brackets as it 
%appears in the second derivative of the length function given in \eq{Lpp}.

The Jacobian ${\cal J}_{v,w}(r)$ becomes zero, and therefore the amplitude 
${\cal A}_{v,w}$ defined by \eq{amp} diverges, at three types of critical points:
 
\begin{itemize}

\item[1)] $r=0$: this corresponds to the caustic point of the families
of degenerate orbits. Making $r>0$ leads to U(1) {\it symmetry breaking}.

\item[2a)] $\cos\beta=0$: this corresponds to a point for which the starting 
and ending angle is $\beta=\pi/2$; the orbit then becomes periodic with $v$ 
reflections. This point is the caustic point $r_{v,w}^{\rm PO}=R\cos\phi_{v,w}$ 
of the PO family $(v,w,1$, an isolated member of which is created there in a 
{\it pitchfork bifurcation} (see \sec{secnlin}).

\item[2b)] The term in square brackets is zero, implying 
\be
v\,r\cos\beta=-R\cos\alpha\,.
\label{bifp}
\ee 
This is fulfilled in the following cases (cf.\ \sec{secnlin}):\\ 
a) for the radial type ``+'' orbits ($\beta=\pi$, $\alpha=0$) with
$(v,w)=(2k+1,k)$, $k>0$, which we call $L_+^{(k)}$ (cf.\ \sec{seclin})
in the points $r_k=R/v=R/(2k+1)$ at which non-radial NPOs $(2k+1,k)$ 
are created in a {\it pitchfork bifurcation}, or\\ 
b) for non-radial orbits which are born in pairs $(v,w)$, $(v,w)'$ in a 
{\it tangent bifurcation} at a critical point $r_{v,w}$ which corresponds 
to a minimum of the function $r(\alpha)$ in \eq{rofang}. This is found 
from the slope function
\be
r'(\alpha)=\frac{{\rm d}r(\alpha)}{{\rm d}\alpha}=\frac{R}{\sin^2\beta}\,
           [\cos\alpha\sin\beta+v\sin\alpha\cos\beta\,]\,.
\label{minrofal}
\ee
The condition $r'(\alpha=0)$ leads, indeed, to \eq{bifp}. 

\item[3)] $r_0=R$: this happens exclusively for the primitive ``+'' orbit 
(1,0) which is responsible for the Friedel oscillations near the surface \cite{rbkm}.

\end{itemize}

\noindent
The regularizations of the diverging amplitudes, using standard uniform approximations
for symmetry breaking and bifurcations, will be discussed in \sec{secunifo}.

The momentum mismatch needed to obtain the density $\delta\tau_1(r)$ 
in \eq{dtau1sc} becomes
\be
Q_{v,w}(r) = {\bf p}\cdot{\bf p'}/p^2 = - \cos(2\beta)\,.
\label{Qvw}
\ee

We stress that the results \eq{Lvw}, \eq{Jvw} and \eq{Qvw} also hold for the
radial orbits with $\alpha=0$ and $\beta=0$ or $\pi$, which are discussed 
specifically in \sec{seclin}, and for which we have the relation $v=2k+1$.

\newpage

\subsection{Periodic orbits}
\label{secPOs}

For $\beta'=\pi-\beta$, we obtain POs. At any point
$r>0$, they are isolated. If $\beta'=\beta=\pi/2$, they are symmetric 
about the radius vector; if $\beta\neq\pi/2$, they exist in
two degenerate versions related to each other by reflection
at the radius vector. Each of them is doubly degenerate due
to time reversal symmetry, except for the diameter orbit $v=2$.

Inserting $\beta'=\pi-\beta$ into \eq{betal}, the angle $\beta$
falls out and we obtain the reflection angle $\alpha_{v,w}$ of
the POs:
\be
\alpha = \alpha_{v,w} = \left(\frac{v-2w}{2v}\right)\!\pi=\frac{\pi}{2}-\frac{w\pi}{v}\,,
\ee
and hence
\be
\phi_{v,w} = \frac{\pi}{2}-\alpha_{v,w} = \frac{w\pi}{v}\,,
\label{phivw}
\ee
as given by Balian and Bloch \cite{babl}. The lengths of the orbits
which are, of course, independent of the starting point $r$, become
\be
L_{v,w}^{\rm PO} = 2vR\,\sin\phi_{v,w}\,.
\label{lpo}
\ee
Note that this formula follows from \eq{Lvw} with \eq{phivw} and $\cos\beta=0$.

For the calculation of the Jacobians, we again use equation \eq{deladelb}, 
but the equation \eq{delrp} becomes for POs, with 
$\sin\beta'=\sin\beta$ and $\cos\beta'=-\cos\beta$,
\be
\delta r'  =
\frac{R}{\sin\beta}\,(\cos\alpha\,\delta\alpha+\sin\alpha\,\cot\beta\,\delta\beta')\,,
\label{delrpp}
\ee
where $\delta\beta'$ is given in \eq{delbetp}. With this, we obtain the Jacobians
\be\hspace{-1.5cm}
{\cal J}_{v,w}(r)^{\rm PO} = -\frac{2v}{Rp_\lambda}\,\frac{1}{\cos\alpha}
                              \left(r^2-R^2\sin^2\!\alpha\right)
                           = -\frac{2v}{Rp_\lambda}\,
                              \frac{\left(r^2-R^2\cos^2\!\phi_{v,w}\right)}{\sin\phi_{v,w}}\,.
\label{jacopo}
\ee
The negative sign in front of the last expression above is opposite to that
given in \cite{disk}, where a trace formula for the level density of the
circular billiard has been derived. It was, however, compensated in \cite{disk}  
by a different book-keeping of the phases in the computation of the Maslov index
[see our remark after equation \eq{mupo} below in \sec{secmors}.] 
Comparing \eq{jnpoR} and \eq{jacopo}, we note that 
${\cal J}_{v,w}^{\mbox{{\scriptsize NPO}}}(r=R)= 
-{\cal J}_{v+1,w}^{\mbox{{\scriptsize PO}}}(r=R)$.
At the caustic points $r_{v,w}^{\rm PO}$ given by 
\be
r_{v,w}^{\rm PO} = R\cos\phi_{v,w}\,,
\label{caust}
\ee
the Jacobian becomes zero. These are, in fact, the critical points at which 
all non-radial POs $(v,w)$ are created from NPOs $(v,w)$ by a pitchfork 
bifurcation (see \sec{secnlin}).

Trivially, the momentum mismatch is $Q_{v,w}(r)=1$ for all POs at all points
except the reflections points on the boundary. At these, $Q_{v,w}(R)$ can
be evaluated as the limit $r\to R$ of the expression \eq{Qvw} for the
(suitably chosen) NPOs.

\subsection{Radial orbits and bifurcations of POs}
\label{seclin}

There are only two kinds of radial NPOs starting at any point $0<r<R$
in the radial direction. They correspond to the only NPOs that exist
in one-dimensional systems. Like in \cite{rb}, we call the ``+''
orbits those which start in the outwards direction and are reflected
at the nearest turning point, and the ``$-$'' orbits those which start
in the inwards direction and are reflected at the opposite turning point. 
The primitive ``+'' and ``$-$'' orbits have only $v=1$
reflection at the boundary; by definition their repetition number
is $k=0$. Using the geometry of \sec{secNPOs}, we have $\alpha=0$ and 
$\beta=\pi$ or $\beta=0$ for the ``+'' or ``$-$'' orbit, respectively. 
To both orbits, one may add $k>0$ full librations between the two 
opposite turning points; we call $k$ their ``repetition number''. 
Their reflection number then is $v=2k+1$. In order to be consistent 
with \eq{betanpo}, we define their winding numbers to be $w=k$ for the 
``+'' and $w=k+1$ for ``$-$'' orbits. Because both $v$ and $w$ are 
determined by $k$, the repetition number $k$ is sufficient to 
characterize the radial NPOs uniquely. We shall denote them here by the 
symbol L$_\pm^{(k)}$. At $r=R$, the orbits L$_+^{(k)}$ with $k>0$ become 
identical with the $k$-th repetitions of the primitive PO (2,1), while 
the orbits L$_-^{(k)}$ for all $k$ become its $(k+1)$-th repetitions $(2k+2,k+1)$.

The formulae \eq{Lvw}, \eq{Jvw} and \eq{Qvw} can be used with 
$\cos\beta=\mp 1$ for the L$_\pm^{(k)}$ orbits. One of their 
characteristic features is that they have opposite initial and final 
momenta, $\bfp'=-\bfp$, so that $Q_\pm^{(k)}(r)=-1$ at all points. Their 
lengths become
\be
L_\pm^{(k)}(r) = 2[(2k+1)R \mp r]\,.
\label{Lpm}
\ee
For the Jacobians we obtain
\be
{\cal J}_\pm^{(k)}(r) = \frac{2}{Rp_\lambda}\,r[(2k+1)r\mp R]\,. 
\label{Jpm}
\ee
Both Jacobians are zero at $r=0$ where the ``+'' and ``$-$'' orbits become 
identical and form a rotationally degenerate family, as discussed in more 
detail in \sec{secsymbr}. Furthermore, the L$_+^{(k)}$ orbits have zero 
Jacobians at $r_k=R/(2k+1)$. For the primitive ``+'' orbit, this gives the 
turning point $r_0=R$ at the boundary, where the divergence of the 
semiclassical amplitude is removed by the uniform approximation
discussed in \sec{secfried}. At the points $r_k=R/(2k+1)$ with $k>0$, 
{\it non-radial} NPOs with $2k+1$ reflections at the boundary are born 
from the L$_+^{(k)}$ orbits in pitchfork bifurcations. The example 
of $k=1$, where the new NPO here is called the $\Lambda$ orbit (see also
\fig{bifs}), will be discussed explicitly in \sec{secunifo} along with 
the corresponding uniform approximation for its semiclassical amplitude.

\subsection{Creation and systematics of non-radial orbits}
\label{secnlin}

The properties of all closed orbits in the circular disk are obtained 
by systematically inverting \eq{rofal} for all $(v,w)$ and inserting
the resulting angles $\alpha(r)$, $\beta(r)$ at each point $r$ into 
the equations \eq{Lvw}, \eq{Jvw} and \eq{Qvw}. As already mentioned,
bifurcations of the orbits occur under variation
of $r$, at which new orbits are created. Furthermore, the U(1) 
symmetry of the continuously degenerate orbit families existing at 
$r=0$ is broken when $r$ becomes $>0$, whereby non-radial NPOs
are created. In fact, we find that all non-radial NPOs are 
created either $(i)$ by the $k$-th repetition of the diametrical 
PO $(v,w)=(2k,k)$ (with $k=1,2\dots$) by U(1) {\it symmetry breaking}, 
or $(ii)$ by the orbits L$_+^{(k)}$ by {\it pitchfork bifurcations}, 
or $(iii)$ pairwise by {\it tangent bifurcations}. All non-radial POs 
are created by pitchfork bifurcations from non-radial NPOs.

The detailed systematics are as follows. (We include here also the Morse
indices whose determination is discussed in \sec{secmors}.)

\begin{itemize}

\item{Non-radial orbits have $v\geq 2$ and $1\leq w \leq 
      \lfloor\frac{v}{2}\rfloor$, where $\lfloor x \rfloor $ is the integer part of $x$.}

\item{NPOs with {\it even $v$ and maximum $w$}, i.e.\ $(v,w)=(2k,k)$,
      are created from the family of POs $(2k,k)$ (i.e., the $k$-th 
      repetitions of the diameter orbit) by U(1) {\it symmetry breaking} 
      at $r=0$ which is the caustic point of the family. These new NPOs 
      exist at all distances $0\leq r\leq R$. Their Morse index is $\mu=3v$.
      At $r=R$ they become equal to the POs $(2k+1,k)$.} Analytical
      expressions for the orbit (2,1), which we here denote by $\nabla$,
      are given in \ref{secdel2}.

\item{NPOs with {\it odd $v$ and maximum $w$}, i.e.\ $(v,w)=(2k+1,k)$,
      are created from the radial L$_+^{(k)}$ orbits by a
      {\it pitchfork bifurcation} at $r_k=R/(2k+1)$ which are 
      critical points of ${\cal J}_+^{(k)}(r)$. They exist at all distances 
      $r_k\leq r \leq R$; at $r=R$ they become equal to the POs $(2k+2,k)$. 
      At $r=r_{v,w}^{\rm PO}$ they create the PO $(2k+1,k)$ in a further 
      {\it pitchfork bifurcation}, whereby $r_{v,w}^{\rm PO}$ is the caustic
      radius \eq{caust} of that PO family. The Morse index is $\mu=6k+2$ for
      $r_k < r < r_{v,w}^{\rm PO}$ and $\mu = 6n+3$ for $r_{v,w}^{\rm PO}
      < r < R$.}

\item{All remaining NPOs are created in pairs $(v,w)$, $(v,w)'$ (with
      identical values of $v$ and $w$) by {\it tangent bifurcations} at 
      critical points $r_{v,w}$ (with $0<r_{v,w}<R$) where the Jacobian 
      ${\cal J}_{v,w}$ becomes zero [cf. equations \eq{minrofal}, \eq{bifp} 
      and \eq{Jvw}]. They exist only for $r_{v,w}\leq r\leq R$. The points 
      $r_{v,w}$ occur as minima of the function $r(\alpha)$ in \eq{rofal}.
      The two orbits correspond to the two branches of the inverse function 
      $\alpha(r)$, whereby the orbit $(v,w)$ has the slope $r'(\alpha)>0$ 
      and the orbit $(v,w)'$ has the slope $r'(\alpha)<0$.
      They have the following properties and behaviors:\\
      -- The orbit $(v,w)$ creates the PO $(v,w)$ by {\it pitchfork
      bifurcation} at the caustic radius $r_{v,w}^{\rm PO}>r_{v,w}$.
      Its Morse index is $\mu=3v-1$ for $r_{v,w}< r < r_{v,w}^{\rm PO}$
      and $\mu=3v$ for $r_{v,w}^{\rm PO} < r < R$.
      At $r=R$, the NPO $(v,w)$ becomes equal to the PO $(v+1,w)$.\\
      -- The orbit $(v,w)'$ undergoes no further bifurcation. Its Morse 
      index is $\mu=3v-2$ at all radii $r_{v,w}< r < R$. At $r=R$,
      the NPO $(v,w)'$ becomes equal to the PO $(v-1,w)$.}

\end{itemize} 
\begin{figure}[t]
\begin{center}
\begin{minipage}{1.0\linewidth}
\hspace{1.8cm}
\includegraphics[width=.9\columnwidth,clip=true]{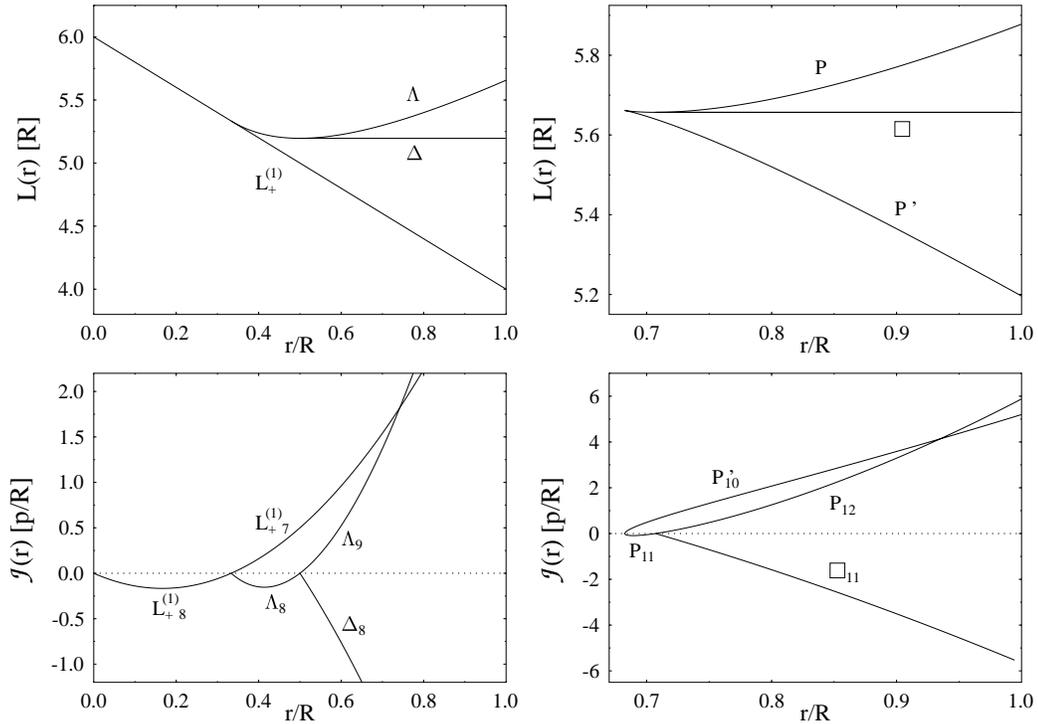}
\end{minipage}
\end{center}\vspace*{-0.5cm}
\caption{
Bifurcations of the NPOs $(v,w)=(3,1)$ (L$_+^{(1)}$ and
$\Lambda$) and the triangular PO (3,1) ($\Delta$) (left panels), 
and of the NPOs $(v,w)=(4,1)$ (P), (4,1)' (P') and the square 
PO (4,1) ($\square$) (right panels). {\it Top:} lengths $L_{v,w}(r)$, 
{\it bottom:} Jacobians ${\cal J}_{v,w}(r)$. In the lower panels,
the subscripts of the orbit symbols indicate their Morse indices.
\label{bifs}}
\end{figure}

In \fig{bifs} we illustrate some of the bifurcation scenarios. In 
the upper panels we show the lengths $L_{v,w}(r)$ and in the lower 
panels the Jacobians ${\cal J}_{v,w}(r)$ of some orbits as
functions of the starting point $r$. In the left panels, the radial
orbit L$_+^{(1)}$ is seen to undergo a pitchfork bifurcation at
$r=R/3$, at which the non-radial orbit $\Lambda$ $(v,w)=(3,1)$ is
created. At $r=R/2$, the orbit $\Lambda$ undergoes a pitchfork 
bifurcation at which the triangular PO $(3,1)$ (noted as $\Delta$) is
born. In the right panels, we see the creation of a pair of orbits 
P (4,1) and P' (4,1)' by a tangent bifurcation at $r_{4,1}=
0.682489R$, to the left of which they do not exist. The orbit P
undergoes a pitchfork bifurcation at $r_{4,1}^{\rm PO}=R/\!\sqrt{2}$,
at which the squared PO (4,1) (noted as $\square$) is born. In the 
lower panels, the subscripts of the orbit symbols indicate their 
Morse indices $\mu$.
\begin{figure}[t]
\begin{center}
\begin{minipage}{1.\linewidth}
\hspace{4.5cm}
\includegraphics[width=0.15\columnwidth,clip=true]{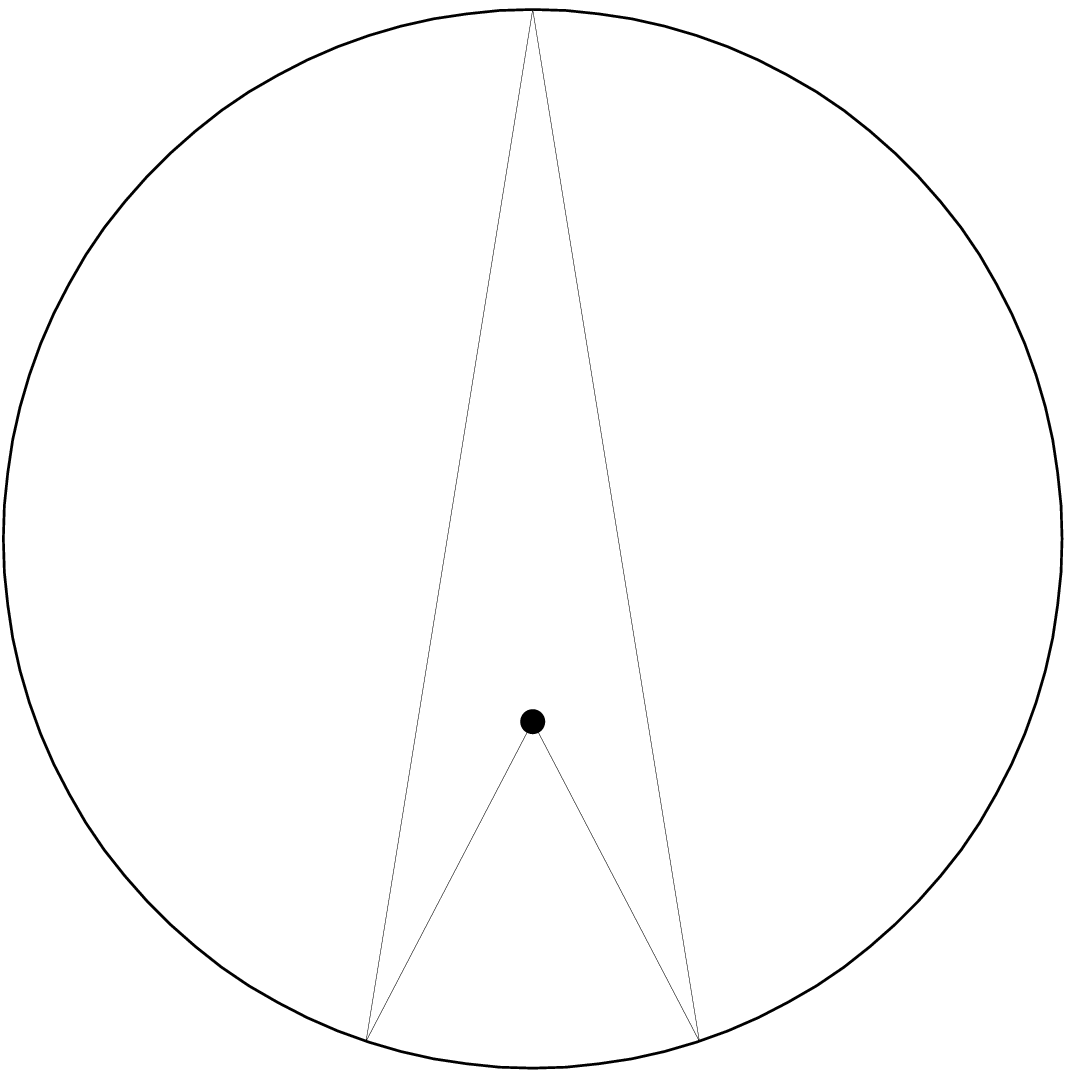}\hspace*{0.8cm}
\includegraphics[width=0.15\columnwidth,clip=true]{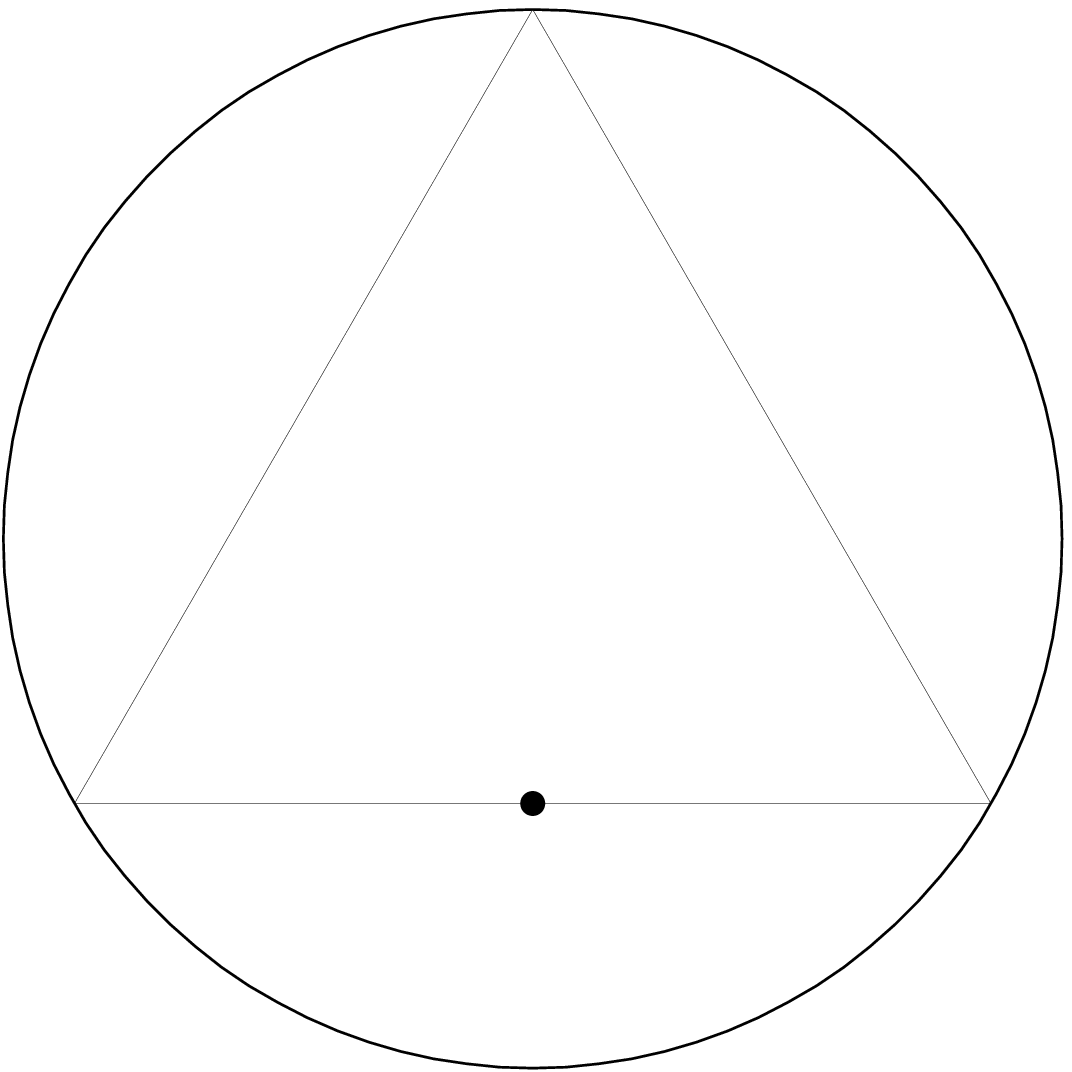}\hspace*{0.8cm}
\includegraphics[width=0.15\columnwidth,clip=true]{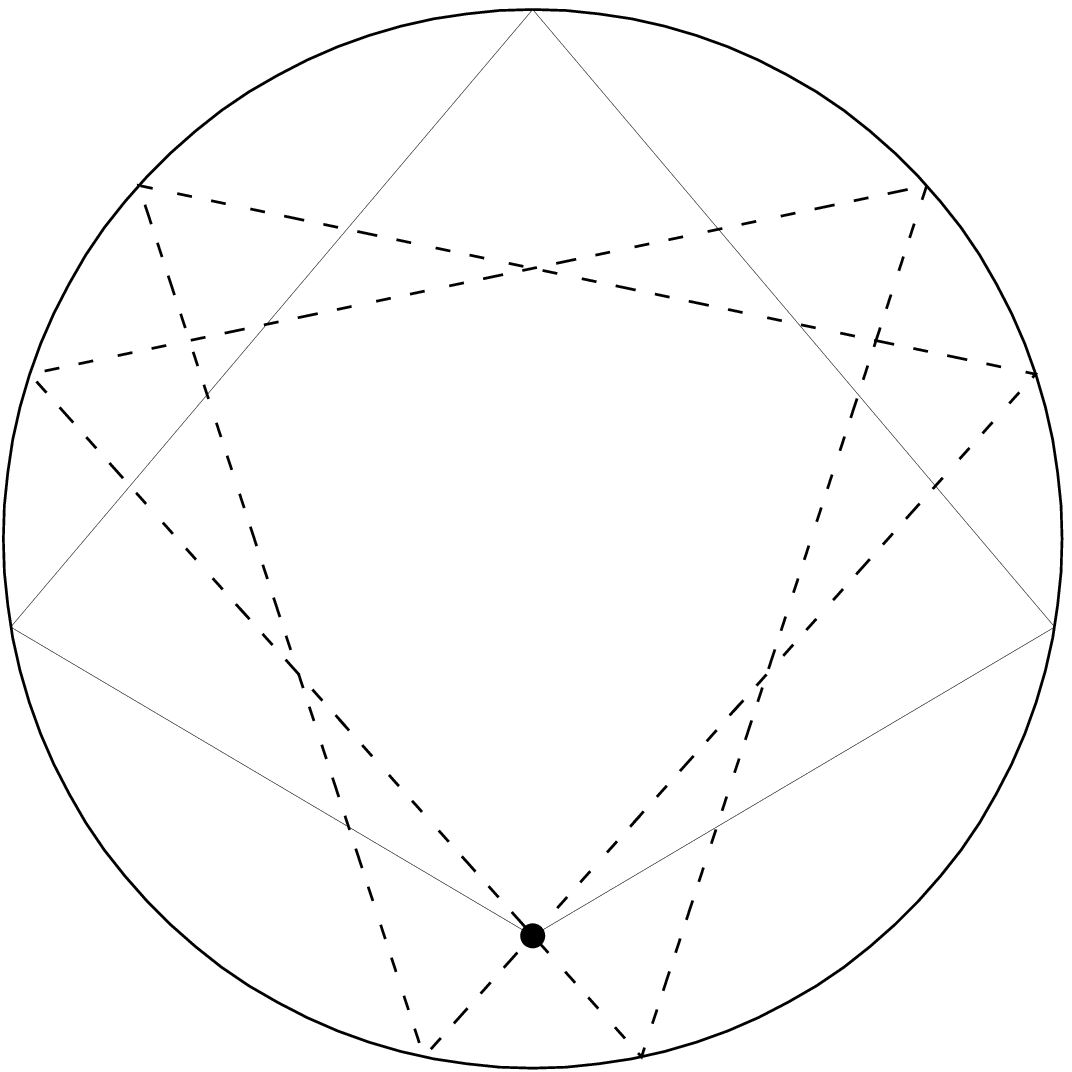}
\end{minipage}
\end{center}
\caption{Shapes of the NPO $\Lambda$ (3,1), shown by the solid lines, at the
starting points $r=0.345R$, $r=R/3$ and $r=0.75R$ (from left to right). In the
right panel, the dashed lines show the shape of the triangle PO
$\Delta$, which exists in two versions lying symmetric to the
diameter that contains the starting point $r$ (black dot).
\label{Lamshapes}
}
\end{figure}
\begin{figure}[t]
\begin{center}
\begin{minipage}{1.\linewidth}
\hspace{4.5cm}
\includegraphics[width=0.15\columnwidth,clip=true]{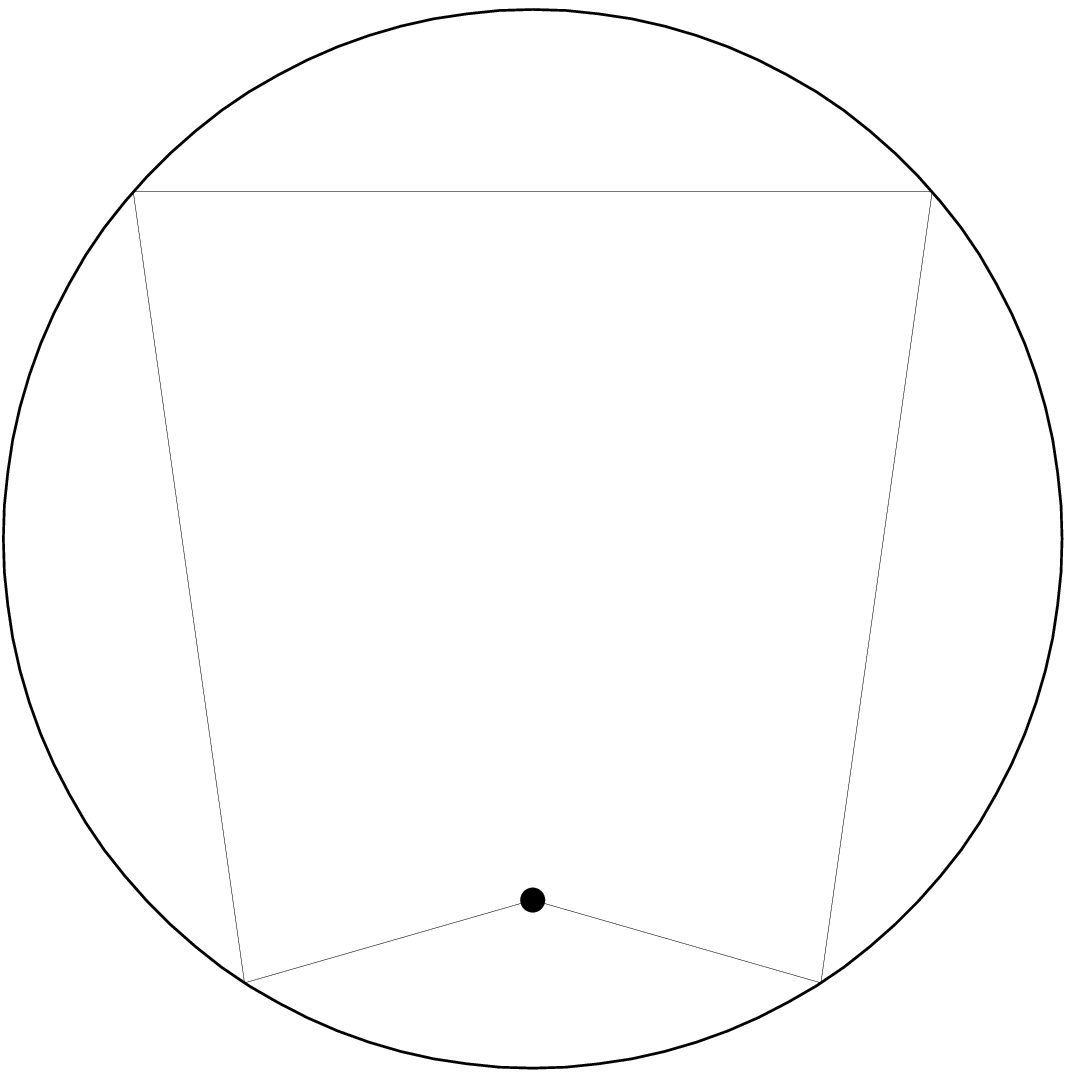}\hspace*{0.8cm}
\includegraphics[width=0.15\columnwidth,clip=true]{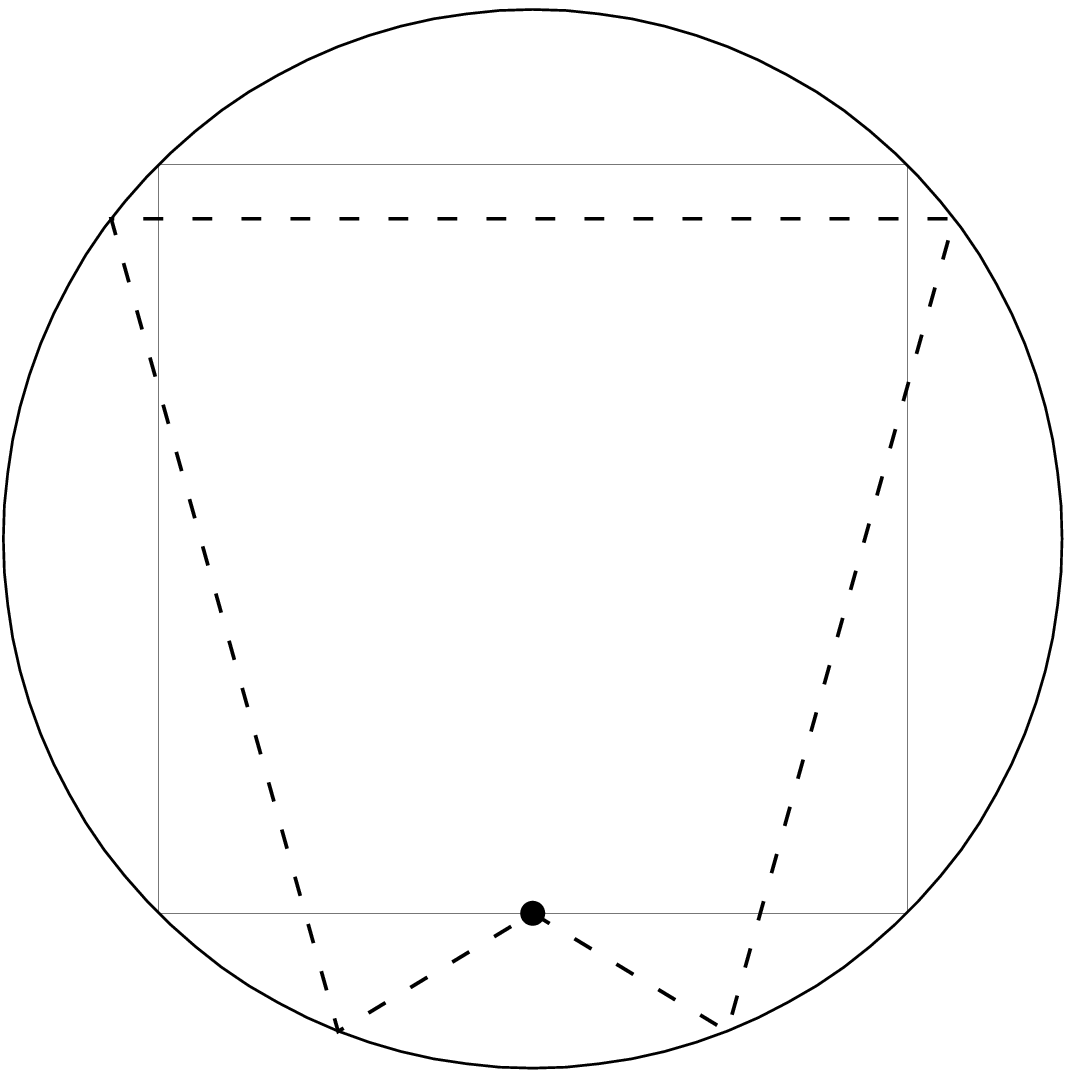}\hspace*{0.8cm}
\includegraphics[width=0.15\columnwidth,clip=true]{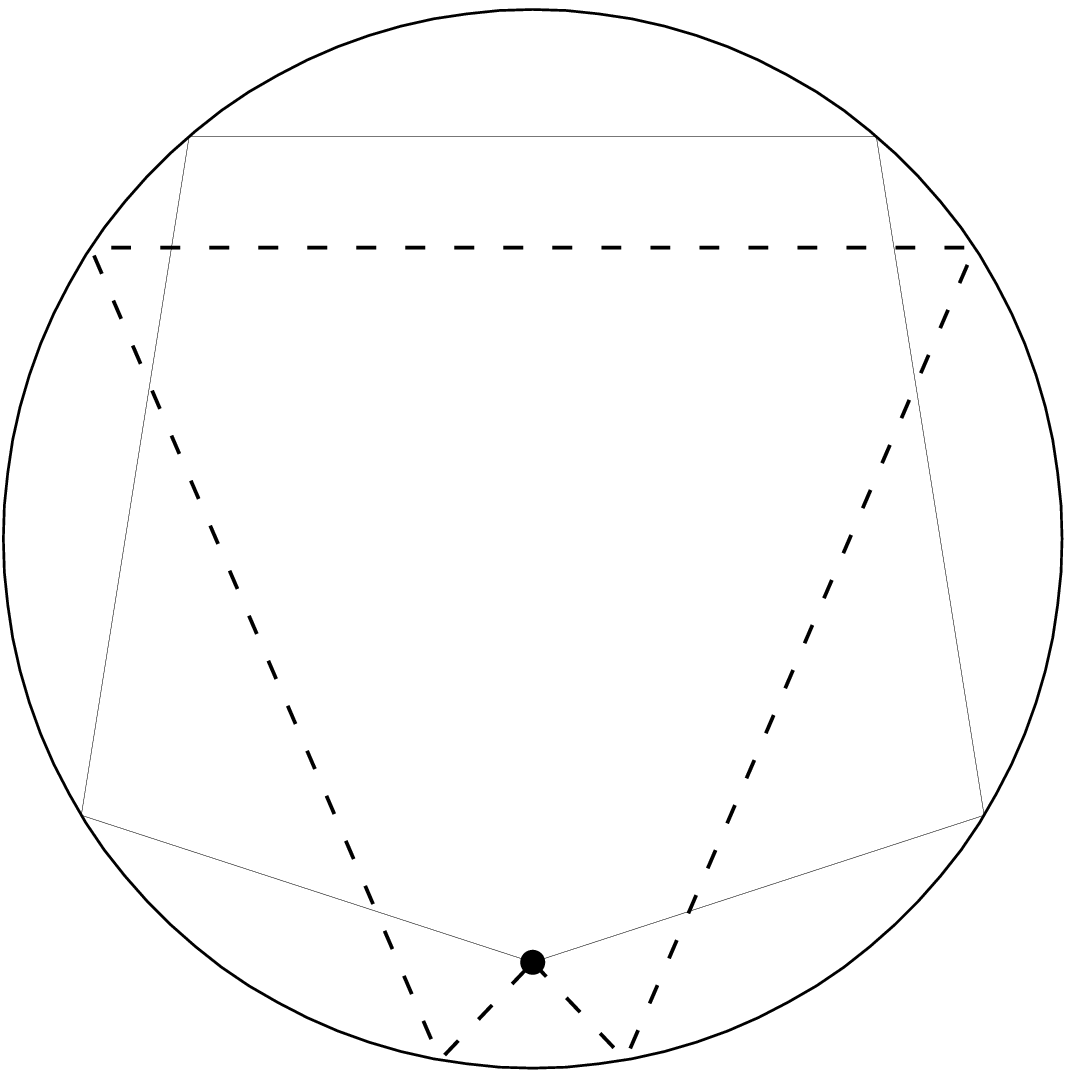}
\end{minipage}
\end{center}
\caption{Shapes of NPOs P (4,1) (solid lines) and P' (4,1)' (dashed lines) 
with starting points $r=r_{4,1}$, $r=R/\!\sqrt{2}$ and $r=0.8R$ (from left to right). 
\label{P4shapes}}
\end{figure}

In \fig{Lamshapes} we show the shape of the $\Lambda$ orbit
at three starting points (from left to right): at $r=0.345R$
shortly after its bifurcation from the L$_+^{(1)}$ orbit,
at $r=R/3$ (i.e.\ at the pitchfork bifurcation where it is 
identical with the PO $\Delta$), and at $r=0.75R$ together 
with the two POs $\Delta$ passing through the same point.
Three shapes of the orbits P and P' are shown in \fig{P4shapes}, 
chosen (from left to right) at the starting point $r=r_{4,1}$ 
(tangent bifurcation, where they are identical), at $r=R/\!\sqrt{2}$ 
(pitchfork bifurcation, where P is identical with $\square$), and at 
$r=0.8R$. Analytical expressions for the lengths and Jacobians of the 
orbits $\Lambda$ and P, P' are given in \ref{seclam3} and \ref{secp41}, 
respectively.
\begin{figure}[h]
\begin{center}
\begin{minipage}{1.\linewidth}
\hspace{0.cm}
\includegraphics[width=0.6\columnwidth,clip=true]{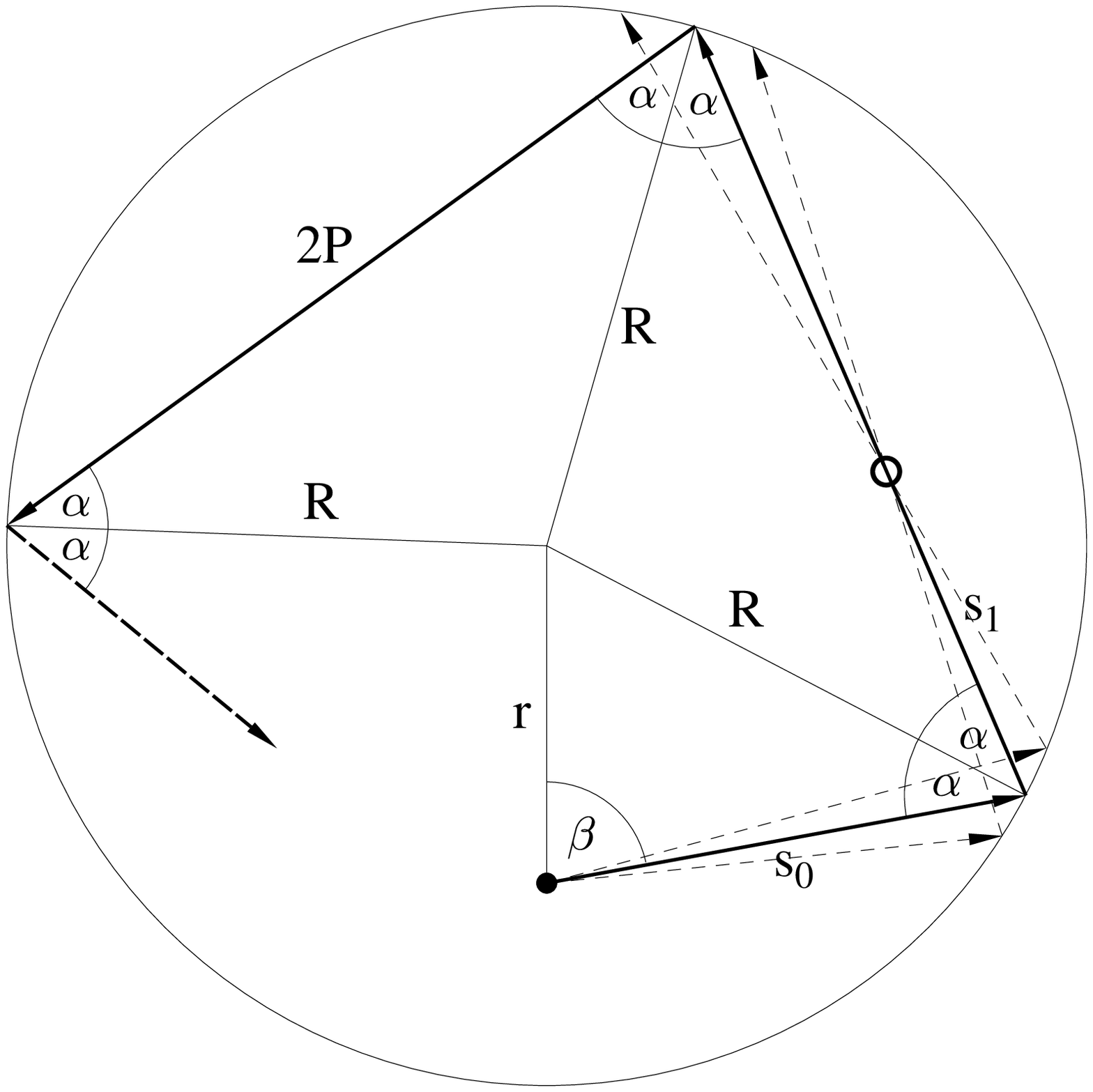}\hspace*{-3.3cm}
\includegraphics[width=0.6\columnwidth,clip=true]{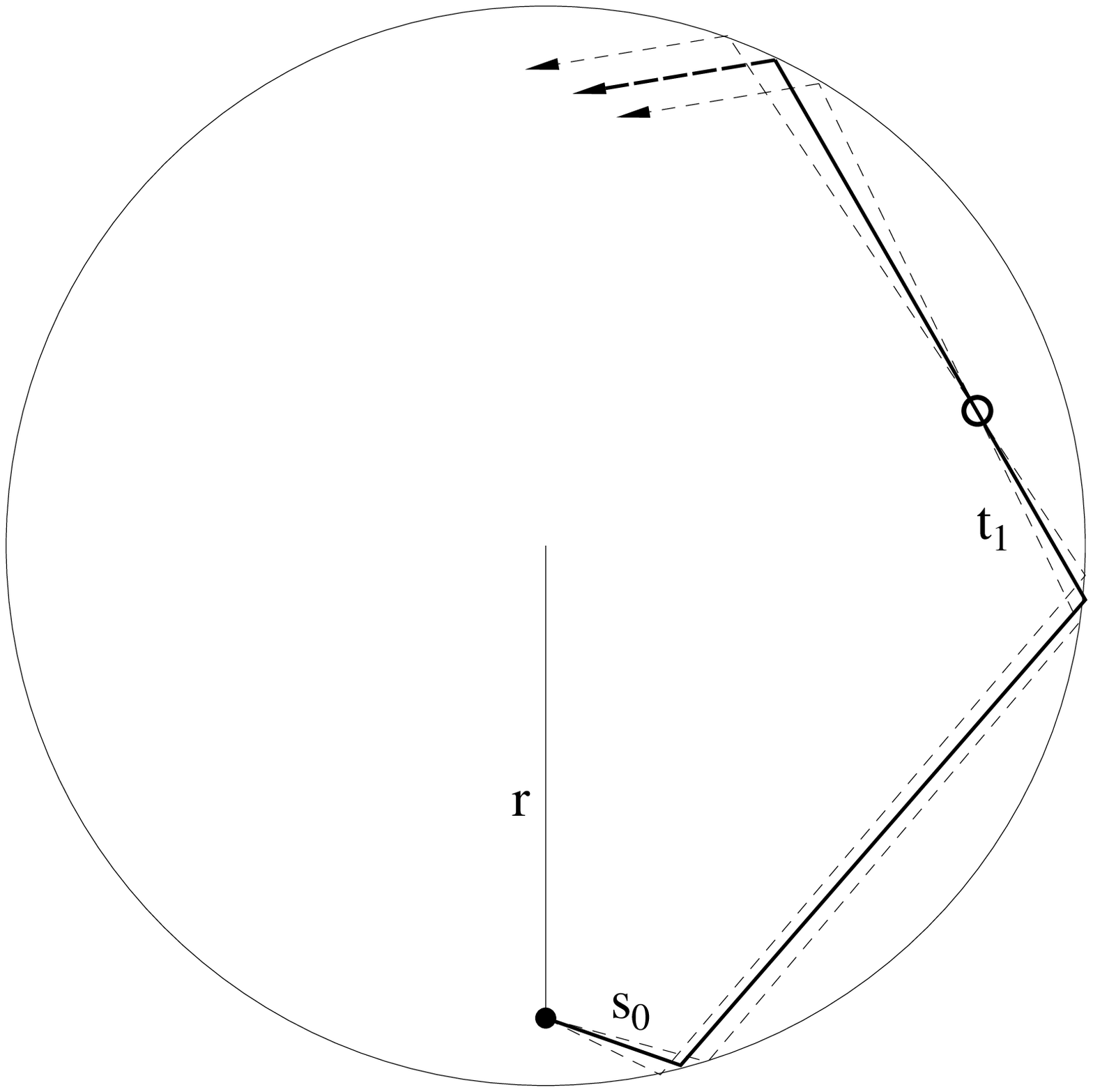}
\end{minipage}
\end{center}\vspace*{-0.5cm}
\caption{
Small perturbations (thin dashed lines) of an orbit (thick solid line)
starting at a point $r$ (black dot) with distance $s_0$ to the first reflection
point. The first conjugate point is shown by a circle. 
{\it Left:} $s_0>2P/3$; here the conjugate point occurs at distance $s_1$ from the
first reflection point. {\it Right:} $s_0<2P/3$; here the conjugate
point occurs only after the second reflection at distance $t_1$ from
the second reflection point.
\label{conjpoints}}
\end{figure}

\subsection{Calculation of Morse indices}
\label{secmors}

Following Gutzwiller \cite{gutz}, we calculate the Morse index of
a given orbit by counting the number of {\it conjugate points} along
the orbit. These are by definition the points $\bfr_{\rm c}$ in which a 
fan of infinitesimally perturbed orbits, all starting at the same 
point $\bfr_0$, intersects itself. To this number, one has to add twice 
the number of reflections at the boundary, since each hard-wall reflection 
yields a phase shift of $\pi$, while each conjugate point that does not 
lie on the boundary yields a semiclassical phase $\pi/2$ \cite{gutz} 
(except for multiple conjugate points which can occur in the presence 
of higher continuous symmetries in $D>2$; this does not happen in the 
present system).

We illustrate the procedure of determining the conjugate points in the
circular billiard in \fig{conjpoints} for an arbitrary non-radial orbit;
the resulting formulae \eq{conjps1}, \eq{conjpt1} hold, however, also 
for the radial NPOs. Let the orbit start 
at a distance $r$ from the center. Let the starting angle be $\beta$, as 
in \fig{circle}, and denote the distance of the starting point to the 
first reflection point by $s_0$. Let the distance of two successive 
reflections of the orbit be $2P$, so that we have
\be
P = R\,\cos\alpha\,.
\label{Pa}
\ee
We now consider a fan of orbits, starting at the same point $r$,
with infinitesimally perturbed angles $\beta'=\beta+\delta\beta$.
These perturbed orbits will intersect in a conjugate point, shown in 
\fig{conjpoints} by a circle. For $s_0 > 2P/3$, the first conjugate 
point is located on the next portion of the trajectory 
at a distance $s_1$ from the first reflection point, as shown in 
the left panel of \fig{conjpoints}. For $s_0 < 2P/3$, the first
conjugate point does not lie on the next portion of the unperturbed 
trajectory, but on the subsequent one, i.e., between the second and 
third reflections, at a distance $t_1$ from the second reflection
point (see the right panel of \fig{conjpoints}).
 
The calculation of the distances $s_1$ and $t_1$ is elementary, using
differential calculus and trigonometry. The perturbed reflection angle 
at the boundary is $\alpha'=\alpha+\delta\alpha$ with
\be
\delta\alpha = \frac{r}{R}\,\frac{\cos\beta}{\cos\alpha}\,\delta\beta\,,
\ee
as follows from \eq{rofang} and \eq{betanpo}. For infinitesimal 
perturbations ($\delta\beta\to 0$), $s_1$ and $t_1$ must be independent 
of $\delta\beta$ due to the definition of a conjugate point. This leads, 
after some algebra using \eq{Pa} and $r\cos\beta=s_0-P$, to the 
following values which are consistent with the findings of \cite{linj}:
\bea
s_1 & = & P - \frac{P(s_0-P)}{(2s_0-P)} \;\qquad \hbox{for} 
              \qquad s_0 \geq \frac23\,P\,, \label{conjps1}\\
t_1 & = & P - \frac{P(P-s_0)}{(3P-4s_0)} \qquad \hbox{for} 
              \qquad s_0 \leq \frac23\,P\,. \label{conjpt1}
\eea
Special cases are the following: ($i$) $s_0=2P/3$ 
$\Rightarrow s_1=2P$ or $t_1=0$. ($ii$) $s_0=0$ 
$\Rightarrow t_1=2P/3$. ($iii$) $s_0=2P$ $\Rightarrow s_1=2P/3$. 
($iv$) $s_0=P$ $\Rightarrow s_1=P$. In this last case, 
the starting point is a {\it caustic point} of a family of degenerate 
orbits due to the rotational U(1) symmetry. We note that these caustic 
points, common to families of rotationally degenerate (periodic or 
non-periodic) orbits, never coincide with conjugate points unless they 
are chosen as starting points (which, however, one should avoid as 
emphasized below).

The successive conjugate points $s_n$ or $t_n$ with 
$n=2,3,\dots$ along an orbit are then found iteratively by taking 
the previous conjugate points as new starting points, evaluating 
their distance $s'_0$ to the next  reflection point, and using the 
relevant formula of \eq{conjps1}, \eq{conjpt1} with $s_0$ replaced 
by $s'_0$.

As stated above, the total Morse index $\mu$ for a given orbit is 
obtained by adding to the number of conjugate points twice the number 
$v$ of reflections at the boundary. We emphasize that for the 
resulting Morse index to be well defined, the starting point may 
{\it not} be any critical (caustic, conjugate or reflection) point
(cf.\ \sec{secregul}). 

For the POs $(v,w)$, we find in this way that the number of conjugate 
points is always $v-1$, i.e., one less than the number of reflections. 
Including the contributions from the reflections, their total Morse 
index therefore is given by
\be
\mu^{\rm PO}_{v,w} = 3v-1\,.
\label{mupo}
\ee
This Morse index has been used implicitly in the derivation of the
trace formula for the level density of the circular billiard in
\cite{disk}. The calculation there was, however, done with a different
book-keeping of the phases, absorbing the negative overall sign of 
${\cal J}_{v,w}$ in our result \eq{jacopo} into the Maslov index 
given in \cite{disk}.

For the radial NPOs L$_\pm^{(k)}$, we obtain the following Morse 
indices:
\bea
\mu_+^{(k)} & = & 6k+2 \quad \hbox{for}\quad r<r_k=R/(2k+1)\,,\nonumber\\ 
\mu_+^{(k)} & = & 6k+1 \quad \hbox{for}\quad r>r_k=R/(2k+1)\,,\nonumber\\ 
\mu_-^{(k)} & = & 6k+3\,.
\label{mulin}
\eea
For the non-radial NPOs, no single general formulae can be found.
We have, however, given the Morse indices of the various types of 
NPOs in their systematics at the end of \sec{secnlin}.

\begin{figure}[t]
\begin{center}
\begin{minipage}{1.\linewidth}
\hspace{4cm}
\includegraphics[width=0.22\columnwidth,clip=true]{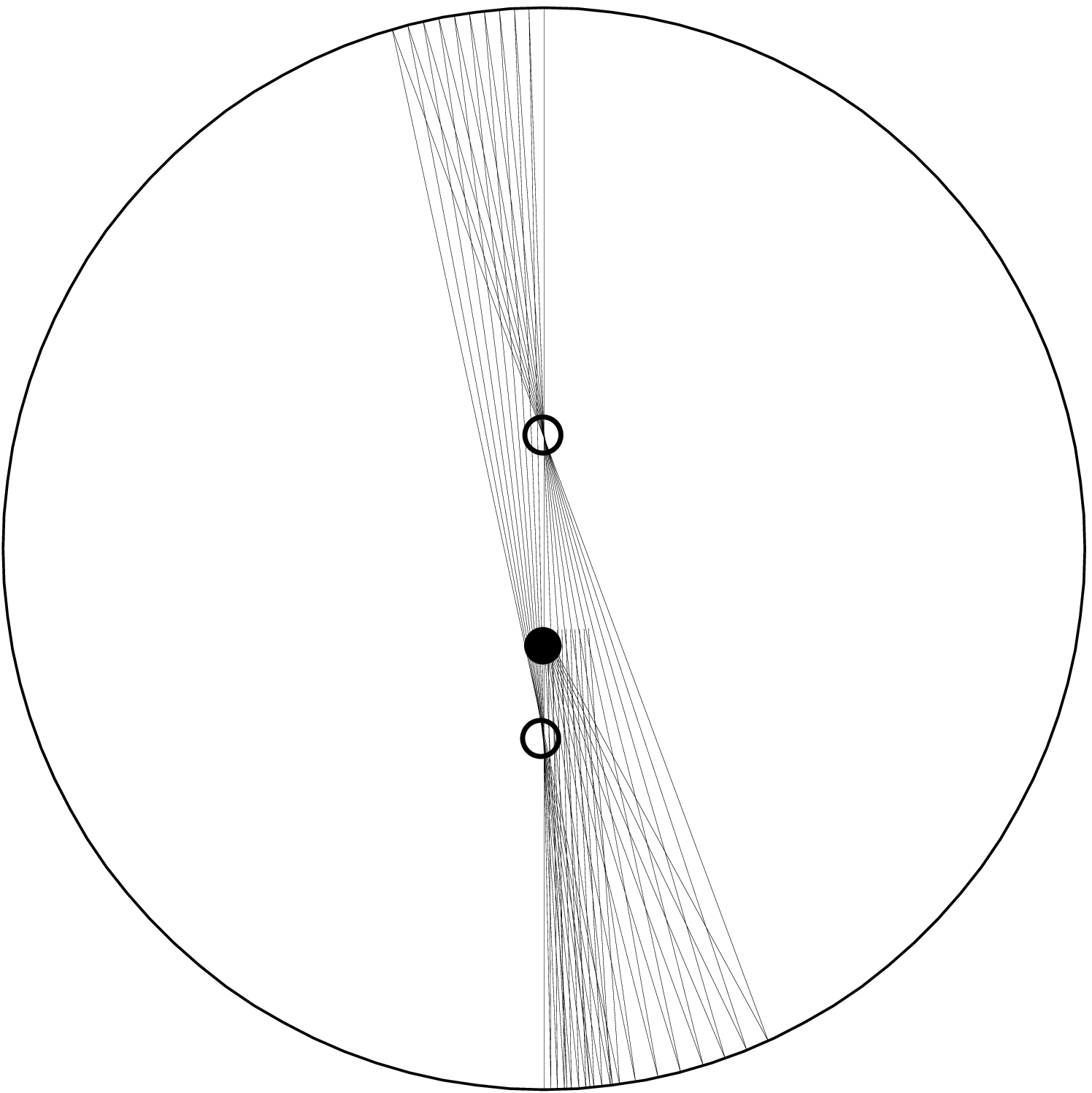}\hspace*{0.5cm}
\includegraphics[width=0.22\columnwidth,clip=true]{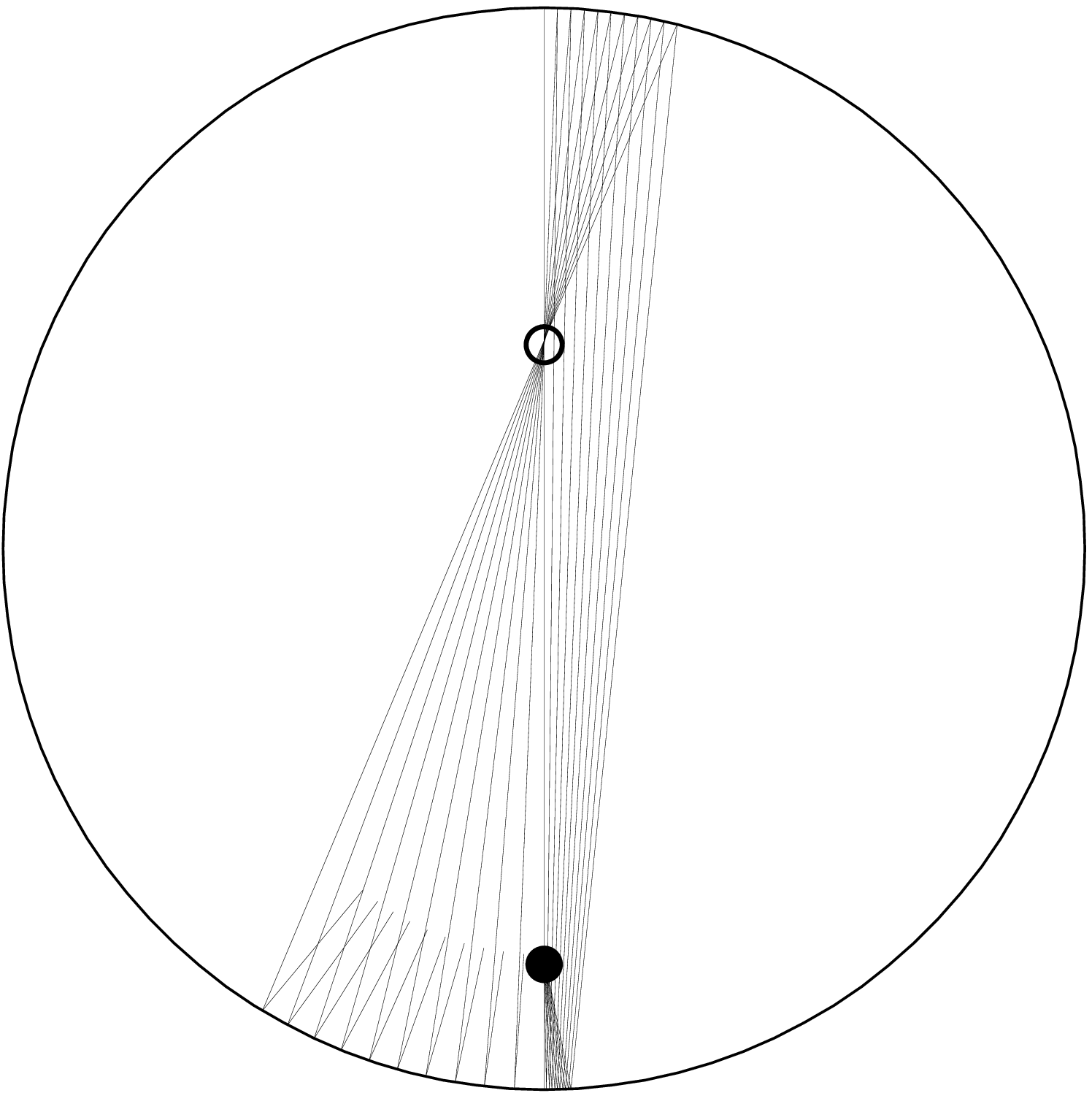}
\end{minipage}
\end{center}
\caption{Orbit L$_+^{(1)}$ with starting point at $r=0.15R$ (left) and 
$r=0.75R$ (right). Fans of small perturbations with different starting 
angles exhibit the positions of the conjugate points. The total Morse 
indices become $\mu=6+2=8$ (left, with two conjugate points) and 
$\mu=6+1=7$ (right, with one conjugate point), in agreement with the
general result \eq{mulin}. 
\label{fig_pls}}
\end{figure}

\begin{figure}[t]
\begin{center}
\begin{minipage}{1.\linewidth}
\hspace{4cm}
\includegraphics[width=0.22\columnwidth,clip=true]{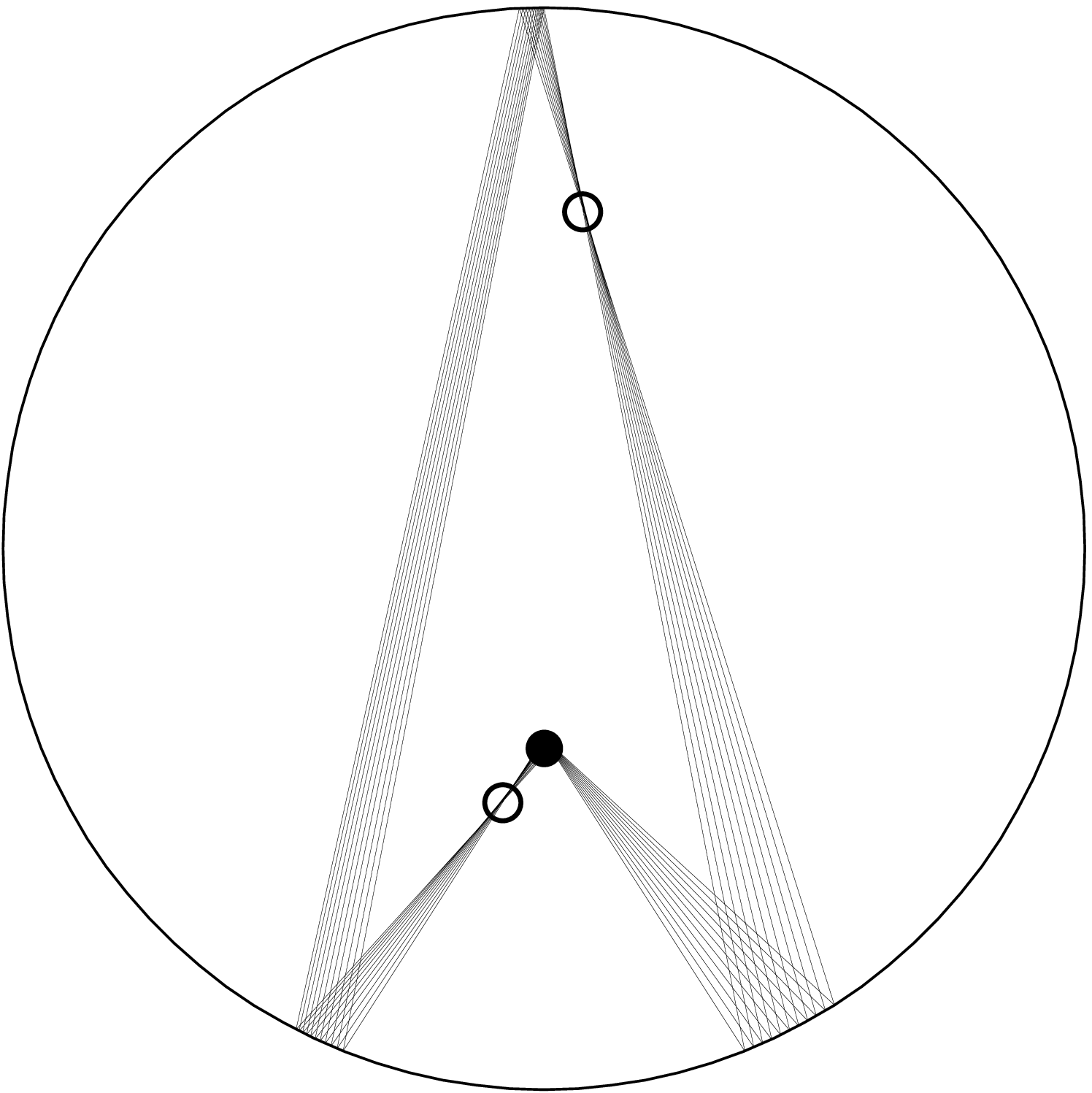}\hspace*{0.5cm}
\includegraphics[width=0.22\columnwidth,clip=true]{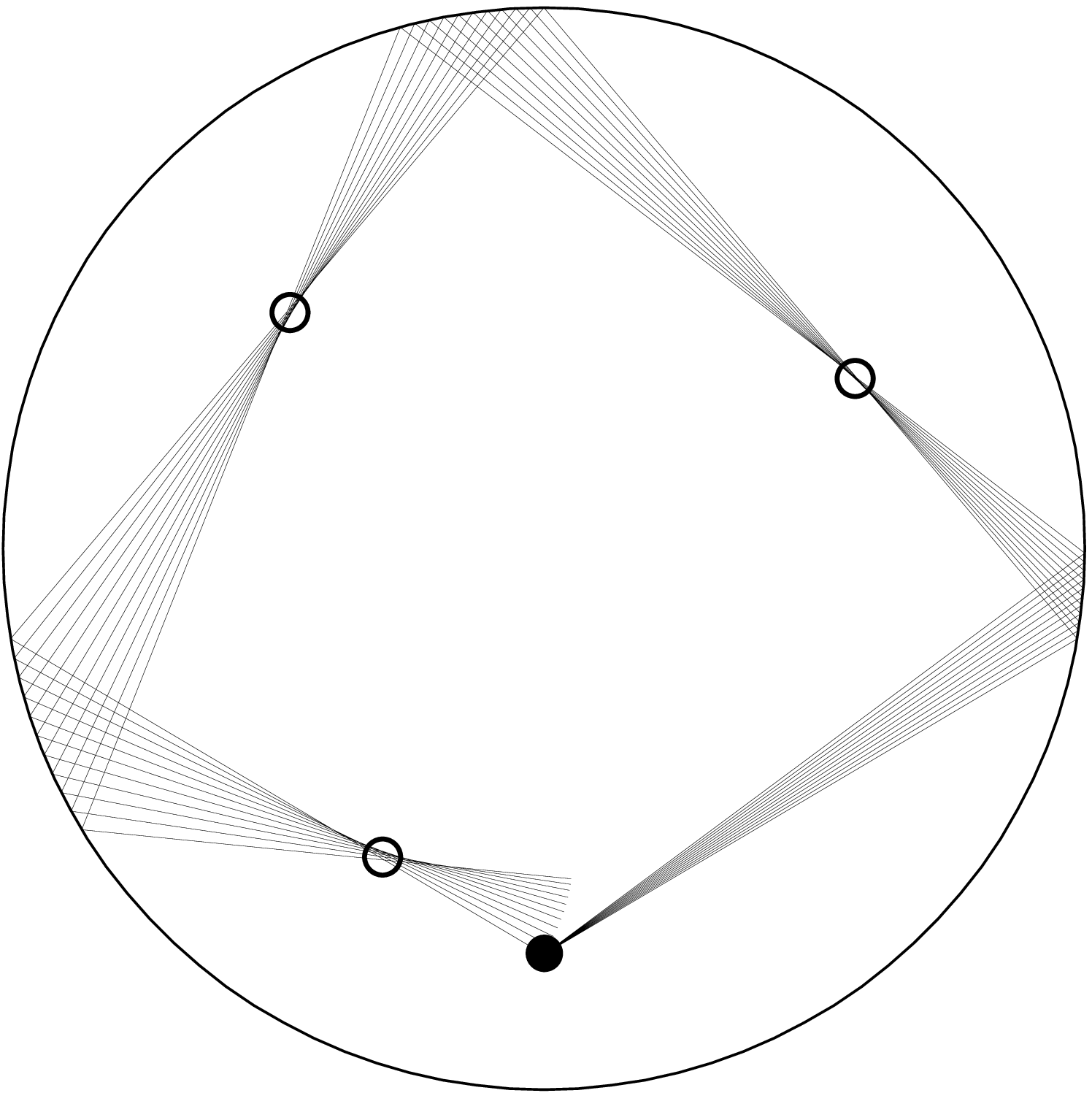}
\end{minipage}
\end{center}
\caption{Same as \fig{fig_pls} for the orbit $\Lambda$ with starting
point at $r=0.35R$, giving $\mu=6+2=8$ (left, with two conjugate points) 
and $r=0.75R$ giving $\mu=6+3=9$ (right, with three conjugate points).
\label{fig_arrow}}
\end{figure}

The geometric construction of the conjugate points for the orbit 
L$_+^{(1)}$ is illustrated in \fig{fig_pls}, and for the orbit $\Lambda$, 
whose bifurcation is shown in \fig{bifs} and further discussed in 
\sec{secunifo}, in \fig{fig_arrow}.

\newpage

\subsection{Relation to multiple-reflection expansion of Green
  function}
\label{secbonche}
In a publication that appears to have remained largely unnoticed by 
the semiclassics community,\footnote{We are grateful to A G Magner for 
drawing our attention to this paper which he discovered recently in the 
preprint collection of the late V M Strutinsky.} Bonche \cite{bonc} has
applied the multi-reflection expansion of the single-particle Green 
function developed by Balian and Bloch \cite{babl} for calculating 
particle densities of three-dimensional billiards. In particular, he
gave explicit results for the spherical billiard whose closed orbits
are identical to those in the two-dimensional circular billiard,
although their semiclassical amplitudes differ since they depend on
the dimension $D$. Bonche discussed the orbits with one reflection,
i.e., the orbits L$_\pm^{(0)}$ in our notation, and obtained similar
results to ours for their regularized contributions at the center and 
the boundary (cf.\ sections \ref{secreglin} and \ref{secfried} below). 
For the orbit with two reflections, i.e., the orbit $\nabla$ (2,1) in 
our notation (cf.\ \ref{secdel2}), he obtained the result 
\eq{xalofr} but gave an obviously wrong expression for its length 
\eq{L2} that would become zero for $r=R$. Bifurcations of NPOs 
have not been noticed in \cite{bonc}.

\newpage

\section{Regularizations of semiclassical amplitudes}
\label{secregul}

At the critical points of ${\cal J}_{v,w}(r)$ discussed in the previous 
section, the semiclassical amplitudes ${\cal A}_\gamma(\lambdab,r)$ in the 
equations \eq{drhosc} -- \eq{dtau1sc} for the spatial densities
diverge. The reason for these divergences can be traced back to a break
down of approximate stat\-ionary-phase integrations used in the derivation 
of Gutzwiller's semiclassical Green function \cite{gutz} that underlies 
also our semiclassical theory of spatial density oscillations \cite{rb}. 
In order to regularize the amplitudes, we resort to standard uniform
approximations that have been developed for semiclassical trace formulae 
for the level density in terms of POs in connection with symmetry breaking 
\cite{ozoh,toms,si96,crpt} and bifurcations \cite{ozoh,si96,ss97,ss98,bt}.

For symmetry breaking, the uniform approximations in general depend not 
only on the particular symmetry that is broken, but also on the way in 
which a system is perturbed \cite{crpt,hhun}. For U(1) symmetry breaking, 
however, the uniform approxi\-mation developed in \cite{toms,si96} is 
universal and can be readily applied in \sec{secsymbr}. 
For bifurcations, on the other hand, the uniform approximations do not
depend on the particular system. They are, in fact, universal for 
each generic type of bifurcation, making use of the standard normal forms 
known from bifurcation theory (see, e.g., \cite{ozob}). It turns out that the
pitchfork and tangent bifurcations occurring for the NPOs in the circular 
billiard have exactly the same generic features as those of POs, so that we 
readily can make use of the corresponding uniform approximations which were 
developed in \cite{ss97}. That for pitchfork bifurcations will be discussed 
in \sec{secunifo}. Finally, the contribution of the primitive ``+''
orbit L$_+^{(0)}$ at $r=R$ to the density $\rho(r)$ can be regularized
as already stated in \cite{rbkm} and briefly illustrated in \sec{secfried}. 
All these regularizations will then be used in our numerical calculations 
in \sec{secnum}. 

We shall only present explicitly the uniform approximations for the density 
oscillations $\delta\rho(r)$, whereby the basic input is given by the 
amplitudes ${\cal A}_\gamma(r)$ defined in \eq{amp}, which
diverge at the critical points. For symmetry breaking and for bifurcations,
the uniform approximations for the kinetic energy densities are then 
easily obtained according to the semiclassical equations \eq{dtausc} 
and \eq{dtau1sc}: For $\delta\tau(r)$, we just have to multiply the
result of $\delta\rho(r)$ by the factor $p_\lambda^2/2m$. For $\tau_1(r)$, 
we must furthermore replace the amplitudes ${\cal A}_\gamma(r)$ 
in the resulting expressions for $\delta\rho(r)$ by 
${\cal A}_\gamma(r) Q_\gamma(r)$. This does, however, not
work for the regularization at the boundary $r=R$, as discussed at
the end of \sec{secfried}.

%\newpage

\subsection{Symmetry breaking at $r=0$}
\label{secsymbr}

\subsubsection{Radial NPOs L$_\pm^{(k)}$.}
\label{secreglin}

The local regularization for the radial orbits L$_\pm^{(k)}$ 
at $r=0$ has already been given in \cite{rb,rbkm} and yields the result
\eq{delrhorad} which is valid for $r\ll R$. Here we first present 
an alternative derivation of this result, following the perturbative 
approach of Creagh \cite{crpt}, using $r/R$ as the perturbation
parameter which for $r>0$ breaks the U(1) symmetry. We then follow
the approach of \cite{toms,si96} to derive a global uniform
approximation that is valid up to all distances $r$ for which no new 
critical phenomenon occurs. (We refer to section 6.3.1 of \cite{book} 
for an easy understanding of the main ideas and the technical details.)

The basic idea is that the NPOs L$_\pm^{(k)}$ at $r>0$ can be
considered as the result of a symmetry-breaking process. At $r=0$, 
they are degenerate, forming a family of orbits with
U(1) symmetry obtained by rotation about an angle $\phi\in[0,2\pi)$. 
For $r>0$ this symmetry is broken; the scenario corresponds to the 
generic case of breaking a U(1) torus into a pair of isolated orbits 
according to the Poincar\'e-Birkhoff theorem \cite{ozob}.

We make use of the stationary property of the length function $L_{v,w}(r)$ in 
\eq{Lvw} at a fixed distance $r$ and for fixed values $v,w$. With the help 
of \eq{rofang} we rewrite it as
\be
L_{v,w}(r,\beta) = 2v\sqrt{R^2-r^2\sin^2\beta}+2r\cos\beta\,.
\label{Lvwofb}
\ee
At the stationary points $\beta_-=0$ and $\beta_+=\pi$, this yields
exactly the lengths of the orbits L$_\pm^{(k)}$ given in \eq{Lpm}
with $v=2k+1$. For $r=0$, the function \eq{Lvwofb} is independent of 
$\beta$ and yields the lengths $2(2k+1)R$ of the degenerate 
orbit families, in short called ``tori''.

We now re-interpret the $L_{v,w}(r,\beta)$ in the following
way. We rename the variable $\beta$ to $\phi$ and take it as the
parameter $\phi\in[0,2\pi]$ describing the U(1) symmetry of the
tori. For $r>0$, $L_{v,w}(r,\phi)$ yields the lengths of the
perturbed tori. To first order in the perturbation parameter 
$r/R$, we thus get the actions of the perturbed tori to be
\be
S^{(k)}(r,\phi) = S_0^{(k)}+2rp_\lambda\cos\phi\,, \qquad 
                  S_0^{(k)}=2p_\lambda(2k+1)R\,.
\label{actphi}
\ee
According to the classical perturbation theory of Creagh \cite{crpt},
we can now write the contribution of the L$_\pm^{(k)}$ orbits to the 
particle density in \eq{drhosc} as the integral
\be
\delta_{\rm r}\rho(r) = {\rm Re} \; \frac{1}{2\pi}\sum_{k=0}^\infty \int_0^{2\pi} \!
                        {\tilde A}_k(r,\phi)\,
                        e^{i[S^{(k)}(r,\phi)/\hbar-\mu_0^{(k)}\pi/2-3\pi/4]}\, {\rm d}\phi\,,
\label{intdens}
\ee 
where $\mu_0^{(k)}$ are the Morse indices of the unperturbed tori which
will be determined below. The amplitude functions ${\tilde A}_k(r,\phi)$ 
will also be determined in the following; they must be chosen such that 
for $r\simeq 0$ the sum of integrals \eq{intdens} leads to the result 
\eq{delrhorad}.

By construction, the {\it stationary points of} $S^{(k)}(r,\phi)$ at
fixed $r$ are $\phi_-=0$ and $\phi_+=\pi$ and yield the actions
of the isolated L$_\pm^{(k)}$ orbits:
\be
S^{(k)}(r,\phi_\pm) = S_\pm^{(k)}(r) = 2p_\lambda [(2k+1)R \mp r]\,.
\label{Spm}
\ee
Hereby $\phi_+$ and $\phi_-$ are the directions into which the 
orbits start from the point $r>0$: the ``+'' orbits start towards 
$\phi_+=\pi$, while the ``$-$'' orbits start to the opposite 
direction $\phi_-=0$. The {\it stationary-phase approximation} for
the integrals \eq{intdens}, with appropriately chosen values
of ${\tilde A}_k(r,\phi)$, therefore yields exactly the
contributions of the isolated L$_\pm^{(k)}$ orbits to \eq{drhosc}. 
Since $S''^{(k)}(r,\phi_+)>0$ and $S''^{(k)}(r,\phi_-)<0$, their 
Morse indices differ by one unit as they must, according to their 
values given in \eq{mulin} for small $r$.

We now take, in a first step, the amplitudes ${\tilde A}_k(r,\phi)$
to be independent of $\phi$, using their forms \eq{amp} valid for 
small $r$, and perform the integral in \eq{intdens} {\it exactly}
rather than by stationary-phase approximation. This leads to the
Bessel function $J_0(2rp_\lambda/\hbar)$ with the correct argument, 
hence reproducing the locally regularized density \eq{delrhorad} after 
summing over all repetitions. This result corresponds to a {\it local}
uniform approximation in the spirit of \cite{ozoh,crpt} which is 
correct locally for $r\simeq 0$ for which the amplitudes were 
approximated. (We recall that using the full amplitudes given by 
\eq{amp}, the summation over $k$ cannot be done analytically.)

In the second step, required to obtain a {\it global} uniform approximation
valid also for large $r$, we choose ${\tilde A}_k(r,\phi)$ to be of the 
form ${\tilde A}_k(r,\phi)=a_k(r)+b_k(r)\cos\phi$. The coefficients 
$a_k(r)$ and $b_k(r)$ are chosen such that they reproduce the {\it exact} 
amplitudes ${\cal A}_\pm^{(k)}(r)$ given in \eq{amp} at large
distances $r$ from the center, where the stationary-phase
approximation for the integral is valid. The exact integral for the 
terms including $b_k(r)$ leads to the Bessel function $J_1(2rp_\lambda)$. 
As a result, we obtain the following global uniform approximation for the 
combined contributions of the radial NPOs L$_\pm^{(k)}$ to the particle 
density oscillations:
\bea
\hspace{-2.cm}\delta_{\rm r}\rho^{\rm (un)}(r)
                   =  \sqrt{4\pi rp_\lambda/\hbar}\; \sum_{k=0}^\infty
                      \left\{{\bar A}_k(r)\,J_0(2rp_\lambda/\hbar)
                      \cos\!\left[\frac{1}{\hbar}S_0^{(k)}-\frac{\pi}{2}\mu_0^{(k)}-\frac{3\pi}{4}\right]
                      \right.\nonumber\\ 
\hspace{2.cm}\left. -\Delta A_k(r)\,J_1(2rp_\lambda/\hbar)
                      \sin\!\left[\frac{1}{\hbar}S_0^{(k)}-\frac{\pi}{2}\mu_0^{(k)}-\frac{3\pi}{4}\right] 
                      \right\}\!,
\label{rhorunifa}
\eea
where
\be
\hspace{-0.8cm}{\bar A}_k(r) = \frac12 \left[{\cal A}_+^{(k)}(r)+{\cal A}_-^{(k)}(r)\right],\quad 
\Delta A(r)   = \frac12 \left[{\cal A}_-^{(k)}(r)-{\cal A}_+^{(k)}(r)\right].
\label{inputun2}
\ee
Hereby ${\cal A}_\pm^{(k)}(r)$ are the exact amplitudes obtained from \eq{amp}
(we have omitted the Fermi energy $\lambdab$ in their arguments),
using the properties of the orbits L$_\pm^{(k)}$ given in \sec{seclin}.

The result \eq{rhorunifa} is finite at $r=0$, where the second term $\propto J_1(2rp_\lambda)$
does not contribute. This term becomes important, however, at larger distances $r$. When 
$2rp_\lambda \gg \hbar$, we may use the asymptotic forms of the Bessel functions for large 
arguments \cite{abro} (which is tantamount to using the stationary-phase
approximation for the above integrals), and then the result
\eq{rhorunifa} goes over into the contributions of the isolated
L$_\pm^{(k)}$ orbits corresponding to \eq{drhosc} 
\bea
\hspace{-1.2cm}
\delta_{\rm r}\rho^{\rm (un)}(r) \longrightarrow\;\; \sum_{k=0}^\infty {\cal A}_+^{(k)}(r)
                                 \cos\left[\frac{1}{\hbar}S_+^{(k)}(r)
                                 -\frac{\pi}{2}(\mu_0^{(k)}-1/2)-\frac{3\pi}{4}\right] \nonumber\\
               \hspace{1.cm}    +\sum_{k=0}^\infty {\cal A}_-^{(k)}(r)
                                 \cos\left[\frac{1}{\hbar}S_-^{(k)}(r)
                                 -\frac{\pi}{2}(\mu_0^{(k)}+1/2)-\frac{3\pi}{4}\right]\!.
\eea 
In order to obtain the correct Morse indices \eq{mulin} (valid below
the bifurcation points $r_k$ of the ``+'' orbits), we now see that we 
have to choose $\mu_0^{(k)}=6k+5/2$, which corresponds to the average 
value of $\mu_+^{(k)}$ and $\mu_-^{(k)}$. With this, the phases can be
simplified to a common sign factor and the result \eq{rhorunifa} becomes
\bea
\hspace{-1.5cm}\delta_{\rm r}\rho^{\rm (un)}(r)
                   =  \sqrt{4\pi rp_\lambda/\hbar}\;\sum_{k=0}^\infty (-1)^k  
                      \left\{{\bar A}_k(r)\,J_0(2rp_\lambda/\hbar)
                      \,\cos [S_0^{(k)}\!/\hbar]
                      \right.\nonumber\\                
\hspace{3.9cm}\left. -\Delta A_k(r)\,J_1(2rp_\lambda/\hbar)
                      \,\sin [S_0^{(k)}\!/\hbar] \right\}\!.
\label{rhorunif}
\eea
The result \eq{rhorunif} can only be used below the critical points 
$r_k=R/(2k+1)$ at which the ``+'' orbits with $k>1$ bifurcate, or
below the boundary $r=R$ where the amplitude for the primitive ``+'' 
orbit with $k=0$ diverges. Near these critical points, we need 
different uniform approximations that will be discussed in
\secs{secunifo} and \ref{secfried}.

\subsubsection{Non-radial NPOs $(2k,k)$.}
\label{secregnlin}

\begin{figure}[t]
\begin{center}
\begin{minipage}{1.\linewidth}
\hspace{6cm}
\includegraphics[width=0.345\columnwidth,clip=true]{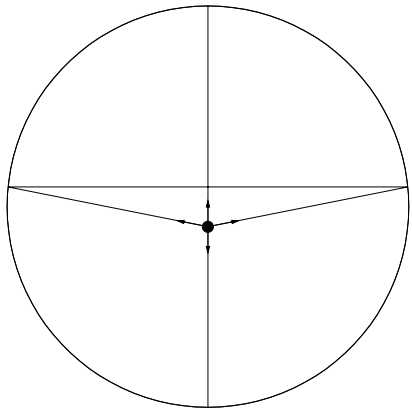}
\end{minipage}
\end{center}
\caption{Non-radial NPO (2,1) (triangle, in the text called $\nabla$) 
starting from a point $r$ (black dot) near the center, where it was 
created from the PO (2,1) (vertical diameter) by U(1) symmetry breaking. 
The arrows indicate the starting directions of the four isolated orbits
(see text for details).
\label{nabla2}
}
\end{figure}

The non-radial NPOs with $(v,w)=(2k,k)$ and $k=1,2,\dots$ are created 
from the $k$-th repetitions $(2k,k)$ of the diametrical PO by symmetry
breaking at $r=0$ (see the systematics in \sec{secnlin}). For small
$r$, they leave their starting points $r$ perpendicular to the
diameter containing the POs $(2k,k)$. As an example, we show in 
\fig{nabla2} the NPO (2,1), which we here call $\nabla$, with a 
starting point $r$ (black dot) close to the center. For $r=0$, these 
NPOs are degenerate with the families of diametrical POs and their 
semiclassical amplitudes diverge. The uniform approximation for these 
NPOs can be derived in a similar way as above for the radial NPOs. 
We first give the derivation for the orbit $\nabla$ (whose analytical 
properties are given in the appendix) and then generalize the result 
to all NPOs $(2k,k)$.
For $r=0$, the PO orbit family $(2,1)$ has the action $S_{\rm 1r}=
4p_\lambda R$. For $r>0$, the family breaks up into a degenerate pair 
of isolated diametrical POs, starting in opposite directions along 
the diameter going through $r$ (this is the vertical diameter in
\fig{nabla2}), plus a discrete pair of NPOs $\nabla$ related by 
time-reversal symmetry (cf.\ the triangle in the figure, starting
in the two directions indicated by the arrows). Their action is
$S_2(r)=p_\lambda L_{2,1}(r)$ with the length $L_{2,1}(r)$ given in
\eq{L2}. We can obtain the actions of all four isolated orbits from 
the stationary points of the action function
\be
S_2(r,\phi) = S_{\rm 1r}+2\Delta {\tilde S}(r)\sin^2\!\phi\,,\qquad
\Delta {\tilde S}(r) = r^2p_\lambda/\!R\,.
\label{S2mod}
\ee
At $\phi=0$ and $\pi$, it yields the action $S_{\rm 1r}$ of the diameter 
orbit, and at $\phi_\pm=\pm\pi/2$, it yields the perturbed action 
$S_{\rm 1r}+2\Delta{\tilde S}(r)$ of the $\nabla$ orbit, valid for small 
$r$ as given in \eq{expLJ2}. Note that the values of $\phi$ at these 
stationary points correspond to the directions (shown by the arrows 
in \fig{nabla2}) in which the perturbed isolated orbits leave the
point $r$ (for the $\nabla$ orbits only in the limit $r\to 0$ where 
the symmetry breaking occurs).

We thus write the common uniform contribution of the four perturbed 
isolated orbits to the particle density \eq{drhosc} as the following integral:
\be
\delta\rho_{2,1}(r) = {\rm Re} \; \frac{1}{2\pi} \int_{-\pi}^{+\pi}
                      {\tilde A}(r,\phi)\,
                      e^{i[S_2(r,\phi)/\hbar-\mu_0\pi/2-3\pi/4]}\,{\rm d}\phi\,.
\label{dr2uni}
\ee
For a constant amplitude ${\tilde A}(r,\phi)$ independent of $\phi$, 
the integral in \eq{dr2uni} yields the Bessel function 
$J_0(\Delta {\tilde S}(r)/\hbar)$, corresponding to a local uniform 
approximation valid near $r=0$. Adding to ${\tilde A}(r,\phi)$ a term 
proportional to $\sin^2\!\phi$ yields an additional Bessel function $J_1$
with the same argument. Proceeding like above for the radial NPOs
and replacing the perturbative action change $\Delta{\tilde S}(r)$ in
the arguments of the Bessel functions by the exact difference
$\Delta S(r)$ given in \eq{actdif2} below, we find the global uniform 
approximation for the combined NPOs and POs (2,1) to be
\bea
\hspace{-2.cm}
\delta\rho_{2,1}^{\rm (un)}(r)
                     = \sqrt{\frac{2\pi\Delta S(r)}{\hbar}}
                       \left\{{\bar A}(r)\,J_0\!\left(\frac{\Delta S(r)}{\hbar}\right)
                       \cos\!\left[\frac{1}{\hbar}{\bar S}(r)
                       -\frac{\pi}{2}\mu_0-\frac{3\pi}{4}\right]\right.\nonumber\\ 
\hspace{2.cm}\left.  +\Delta A(r)\,J_1\!\left(\frac{\Delta S(r)}{\hbar}\right)
                       \sin\!\left[\frac{1}{\hbar}{\bar S}(r)
                       -\frac{\pi}{2}\mu_0-\frac{3\pi}{4}\right] \right\}\!.
\label{rho2unif}
\eea
The amplitudes above are given by
\be\hspace{-1.cm}
{\bar A}(r) = \frac12\, [{\cal A}_{\rm 1r}(r)+{\cal A}_{2,1}(r)]\,,\qquad 
\Delta A(r) = \frac12\, [{\cal A}_{\rm 1r}(r)-{\cal A}_{2,1}(r)]\,,
\label{input2}
\ee
where ${\cal A}_{\rm 1r}(r)$ is the amplitude \eq{amp} evaluated for the diameter
PO (2,1) in term of its properties given in \sec{secPOs} and ${\cal A}_{2,1}(r)$
that of the $\nabla$ orbit using its exact properties given in \eq{L2}, \eq{J2}.
Both amplitudes must include a discrete degeneracy factor 2 corresponding
to their two starting directions. Furthermore, the quantities $\Delta S(r)$ 
and ${\bar S}(r)$ are defined by
\be
\Delta S(r) = \frac12\,[S_{2,1}(r)-S_{\rm 1r}]\,,\qquad 
{\bar S}(r) = \frac12\,[S_{2,1}(r)+S_{\rm 1r}]\,,
\label{actdif2}
\ee
where $S_{2,1}(r)$ is the {\it exact} action of the $\nabla$ orbit, given by 
$S_{2,1}(r)=p_\lambda L_{2,1}(r)$ using \eq{L2}. 

For large enough distances $r$ such that $\Delta S(r)\gg \hbar$, we
can use the asymptotic forms of the Bessel functions for large arguments,
and \eq{rho2unif} goes over into the sum of the contributions of the
isolated diameter POs and $\nabla$ orbits 
\bea
\delta\rho_2^{\rm (un)}(r) \longrightarrow \;
                A_{\rm 1r}(r)\cos\left[\frac{1}{\hbar}
                S_{\rm 1r}-\frac{\pi}{2}(\mu_0-1/2)-3\pi/4\right] \nonumber\\
\hspace{2.3cm}+ A_2(r)\cos\left[\frac{1}{\hbar}S_{2,1}(r)-\frac{\pi}{2}(\mu_0+1/2)-3\pi/4\right]\!.
\label{rho2unifasy}
\eea
With the choice $\mu_0=11/2$, we obtain the correct Morse indices
$\mu_{\rm 1r}=5$ for the diameter orbit and $\mu_{2,1}=6$ for the 
$\nabla$ orbit.

The uniform result \eq{rho2unif} gives a finite combined contribution
of the degenerate orbit families at $r=0$. Since the orbit $\nabla$ 
undergoes no bifurcation and its amplitude stays finite for all $r>0$, 
the uniform result \eq{rho2unif} is valid in the entire region $0\leq r\leq R$.

The result \eq{rho2unif} looks very similar to the uniform
contribution \eq{rhorunif} of the radial NPOs, except that the
argument of the Bessel functions here is quadratic in $r$ (for small
$r$), which yields a slower departure from its maximum. This is
characteristic of the long-ranged irregular oscillations 
$\delta_{\rm irr}\rho(r)$ that are due to non-radial NPOs. In 
fact, the presently discussed class of non-radial NPOs $(2k,k)$ 
is dominating the irregular part of the density oscillations near 
$r=0$, as we shall see in \sec{secnum}.

We therefore now generalize \eq{rho2unif} to include all NPOs of 
type $(2k,k)$ that are created by symmetry breaking at $r=0$ from 
the $k$-fold repetitions of the periodic diameter orbit (2,1). For 
that we simply have to substitute the amplitudes ${\cal A}_{2,1}(r)$ 
and actions $S_{2,1}(r)$ in the equations \eq{input2},
\eq{actdif2} by those obtained from the general formulae \eq{Lvw} 
and \eq{Jvw} for the NPOs with $(v,w)=(2k,k)$, and the quantities 
${\cal A}_{\rm 1r}(r)$ and $S_{\rm 1r}$ by those obtained from
\eq{lpo} and \eq{jacopo} for the POs. The Morse index in \eq{rho2unif} 
must be substituted by $\mu_0^{(k)}=6k-1/2$. After these substitutions 
we must, of course, sum over all $k>0$. Taking out a common phase 
factor, we obtain the global uniform approximation for the dominating 
contribution to the irregular density oscillations:
\bea
\hspace{-2.5cm}
\delta_{\rm irr}\rho^{\rm (un)}(r)
                   =  \sum_{k=1}^\infty \sqrt{\frac{2\pi\Delta S_k(r)}{\hbar}}\,(-1)^k
                      \left\{{\bar A}_k(r)\,J_0\!\left(\frac{\Delta S_k(r)}{\hbar}\right)
                      \cos\!\left[\frac{1}{\hbar}{\bar S}_k(r)
                      +\frac{\pi}{4}\right]\right.\nonumber\\ 
\hspace{2.9cm}\left. +\Delta A_k(r)\,J_1\!\left(\frac{\Delta S_k(r)}{\hbar}\right)
                      \sin\!\left[\frac{1}{\hbar}{\bar S}_k(r)
                      +\frac{\pi}{4}\right] \right\}\!,
\label{rho2kkunif}
\eea
where the amplitudes are
\be
\hspace{-2.cm}{\bar A}_k(r) = \frac12\, [{\cal A}_{k{\rm r}}(r)+{\cal A}_{2k,k}(r)]\,,\qquad 
              \Delta A_k(r) = \frac12\, [{\cal A}_{k{\rm r}}(r)-{\cal A}_{2k,k}(r)]\,,
\label{input2kk}
\ee
and the action functions are
\be
\hspace{-2.cm}
\Delta S_k(r) = \frac12\,[S_{2k,k}(r)-S_{k{\rm r}}]\,,\quad
{\bar S}_k(r) = \frac12\,[S_{2k,k}(r)+S_{k{\rm r}}]\,,\quad
       S_{k{\rm r}} = 4kp_\lambda R\,.
\ee
The subscripts ``$k{\rm r}$'' refer to the diametrical POs with their
properties given in \sec{secPOs}, and the subscripts ``$2k,k$'' to
the NPOs with their properties given in \sec{secNPOs}.

The NPOs $(2k,k)$ exist at all $r>0$ and undergo no bifurcations.
The uniform result \eq{rho2kkunif} is therefore valid in the whole
range $r\in[0,R]$. In the numerical calculations presented in
\sec{secnum}, it will be seen that the summation over $k$ can
be practically limited to $k_{\rm max}\sim 10-20$ in all cases.

\subsection{Uniform approximations for bifurcations}
\label{secunifo}

Here we discuss the uniform approximations needed to
regularize the divergences of the semiclassical amplitudes
arising at bifurcations of the closed orbits. To our knowledge,
there exists no systematic study and classification of 
bifurcations of {\it non-periodic} closed orbits in the literature.
We therefore had to convince ourselves that the bifurcations
occurring in our present system are of the same generic types
as those known \cite{ozob,meye} for periodic orbits. This
appears, indeed, to be the case. For the pitchfork bifurcations,
we shall demonstrate this in a particular example in \sec{secpitch}
and then present the corresponding uniform approximation for
the semiclassical particle density. In \sec{sectang} we briefly
discuss the tangent bifurcations.

\subsubsection{Pitchfork bifurcations.}
\label{secpitch}

In the left panels of \fig{bifs} we have shown two successive
pitchfork bifurcations. We shall take the example at the
critical radius $r=R/3$ where the non-radial orbit $\Lambda$
(3,1) is created from the radial orbit L$_+^{(1)}$. We draw
particular attention to the bifurcation diagram on the lower
left of \fig{bifs}. For isolated periodic orbits of $D=2$ dimensional
systems, the corresponding diagrams are obtained when the trace 
of the stability matrix, Tr\,M$(\epsilon)$ as function of a 
bifurcation parameter $\epsilon$, is considered. We assume a
bifurcation to occur at the point $\epsilon=0$. In the trace 
formula for the level density of such systems \cite{gutz}, 
the quantity Tr\,M$(\epsilon)-2$ appears under a root in the 
denominator of the semiclassical amplitudes, which diverge at 
the bifurcation point where Tr\,M$(0)=2$. Now, it is a characteristic 
feature \cite{ss97,bt} of a generic pitchfork bifurcation that 
the slopes of the functions Tr\,M$(\epsilon)$ of the two orbits 
involved in the bifurcation fulfill the following relation at the 
bifurcation point:
\be
{\rm Tr\,M}'_B(0) = -2\,{\rm Tr\,M}'_A(0)\,.
\label{slopeM}
\ee
Hereby A is the parent orbit which changes its stability at the
bifurcation point and B the daughter orbit that is created at 
the bifurcation and only exists on one side of it. (For a 
mathematical proof of this ``slope theorem'', we refer to 
\cite{jaen1}.)

In our present system, the decisive quantity under a root in
the denominator of the semiclassical amplitudes \eq{amp} is the
Jacobian ${\cal J}_{v,w}(r)$, and $r$ is the bifurcation
parameter. In the example of \fig{bifs} under discussion, 
L$_+^{(1)}$ is the parent orbit A and $\Lambda$ is the
daughter orbit B existing only above the bifurcation point
$r=R/3$. Using the explicit results \eq{Jpm} and \eq{J3}
for the Jacobians of these two orbits, we find
\be \left.
\frac{\rm d}{{\rm d}r}{\cal J}_{3,1}(r)\right|_{r=R/3} = -\frac{4}{p_\lambda}\,,\qquad\left. 
\frac{\rm d}{{\rm d}r}{\cal J}^{(1)}_+(r)\right|_{r=R/3} = \frac{2}{p_\lambda}\,,
\ee 
so that the slope theorem \eq{slopeM} is, indeed, satisfied.

A pitchfork bifurcation which is generic in the sense of Meyer 
\cite{meye} is period-doubling: the new orbit B has twice
the period of the parent orbit A at the bifurcation point.
However, in systems with discrete symmetries the pitchfork 
bifurcations are typically isochronous \cite{bt,then,lame}: 
both orbits have the same periods at the bifurcation point, but
the daughter orbit B has a double discrete degeneracy compared
to the parent orbit A. For the NPOs discussed here, the
pitchfork bifurcation is of the second, non-generic type:
their running times $T(r,\lambdab)$ (given here simply by
their lengths) are identical at the bifurcation point, but 
the $\Lambda$ orbit has a degeneracy of two due its  
time-reversal symmetry while the L$_+^{(1)}$ orbit is not 
degenerate.

Finally, the following rule is known \cite{ss97} for the 
Maslov index appearing in the trace formula: at the pitchfork 
bifurcation, the index of the parent orbit A changes by one
unit, and the index of the orbit B existing after the 
bifurcation is the same as that of the orbit A before the 
bifurcation. Using our results in \sec{seclin} and the
appendix, we see that this rule is, indeed, also fulfilled
here by the Morse indices: the Morse index of the L$_+^{[1)}$ 
orbit changes from 8 to 7 at $r=R/3$, and the $\Lambda$ 
orbit has the Morse index 8 close to $r>R/3$.
After these affirmations of the nature of the present pitchfork 
bifurcation, we can immediately apply the uniform 
approximation that was developed in \cite{ss97} to obtain 
the following common contribution of the L$_+^{(1)}$ and $\Lambda$
orbits to the particle density oscillations in terms of Bessel 
functions of non-integer orders:\footnote{We can also derive 
this result by making 
an ansatz similar to \eq{dr2uni} and expanding the action function
$S_{3,1}(r,\beta)=p_\lambda L_{3,1}(r,\beta)$ using \eq{L3} in 
the phase of the integrand around the stationary point $r=R/3$, 
$\beta=\pi$. Expanding it up to first order in $\epsilon=r-R/3$ 
and fourth order in $\delta=\beta-\pi$, we obtain exactly the 
(one-dimensional) normal form of the action for the pitchfork 
bifurcation that was used in \cite{ss97} to derive the uniform 
approximation}
\bea
\hspace{-2.5cm}
\delta\rho_{\rm pb}^{\rm (un)}(r)
                 =  {\rm Re} \; \sqrt{\frac{\pi\Delta S(r)}{2\hbar}}\,
                      e^{i[{\bar S}(r)/\hbar-\pi\mu_0/2-\pi]}\nonumber\\
\hspace{-0.6cm}\times
                      \left\{ {\bar A}(r)
                      \left[\sigma_0 J_{1/4}\!\left(\frac{\Delta S(r)}{\hbar}\right)
                      e^{i\sigma_1\pi/8}
                      +J_{-1/4}\!\left(\frac{\Delta S(r)}{\hbar}\right)
                      e^{-i\sigma_1\pi/8}\right] \right. \nonumber\\ 
\hspace{-0.6cm}\left. +\Delta A(r)
                      \left[ J_{3/4}\!\left(\frac{\Delta S(r)}{\hbar}\right)
                      e^{i3\sigma_1\pi/8}
                      +\sigma_0 J_{-3/4}\!\left(\frac{\Delta S(r)}{\hbar}\right)
                      e^{-i3\sigma_1\pi/8}\right]\!\right\}\!,
\label{rho3unif}
\eea
with
\bea
\hspace*{-0.8cm}
{\bar A}(r) = \left[\frac{{\cal A}_{3,1}(r)}{2}+\frac{{\cal A}_+^{[1)}(r)}{\sqrt{2}}\right],\qquad
\Delta A(r) = \left[\frac{{\cal A}_{3,1}(r)}{2}-\frac{{\cal A}_+^{[1)}(r)}{\sqrt{2}}\right],\nonumber\\
\hspace*{-0.8cm}{\bar S}(r) = \,\frac12\,[S_{3,1}(r)+S_+^{[1)}(r)]\,,\qquad \;\;
\Delta S(r) = \frac12[S_{3,1}(r)-S_+^{[1)}(r)]\,,\nonumber\\
\hspace*{-0.8cm}
\sigma_0    = \;{\rm sign}(r-R/3)\,,\qquad \qquad\quad\;\;\sigma_1=\,{\rm sign}(\Delta S)\,,
\label{input3}
\eea
in terms of the exact amplitudes ${\cal A}_{3,1}(r)$, ${\cal A}_+^{(1)}(r)$ 
and actions $S_{3,1}(r)$, $S_+^{[1)}(r)$ of the $\Lambda$ and L$_+^{[1)}$ 
orbits, respectively. The choice of $\mu_0=7$ yields the correct
Morse indices of the isolated orbits on either side of the bifurcation. 
Note that for $r<R/3$, there exists an imaginary ``ghost'' orbit
$\Lambda$ which contributes with real action $S_{3,1}(r)$ and amplitude
${\cal A}_{3,1}(r)$ which are just the analytical continuations of these
properties of the real $\Lambda$ orbit, as given in \eq{L3} and
\eq{J3}. Note also that the amplitude ${\cal A}_{3,1}$ must include
the degeneracy factor 2 due to the two time orientations of the
$\Lambda$ orbit.

At the bifurcation point, the result \eq{rho3unif} takes the finite value
\be
\delta\rho_{\rm pb}^{\rm (un2)}(R/3)
                = \frac{1}{\hbar^{3/4}}\frac{9p_\lambda^{3/4}\Gamma(1/4)}{32\pi^2R^{5/4}}
                  \cos\left(\frac{16Rp_\lambda}{3\hbar}-\frac{5\pi}{8}\right).
\ee
The uniform approximation \eq{rho3unif} is again global in the 
sense that it goes over into the sum of the isolated 
contributions of both orbits (including the degeneracy of the
$\Lambda$ orbit) sufficiently far away from the bifurcation,
where the argument of the Bessel functions is large enough so 
that their asymptotic forms can be used.

Having confirmed that all other pitchfork bifurcations occurring
in the circular billiard are of the same type, we can apply the
result \eq{rho3unif} to them, replacing in \eq{input3} the amplitudes 
and actions of the L$_+^{(1)}$ orbit by those of the parent orbit, 
those of the $\Lambda$ orbit by those of the daughter orbit 
(irrespective of whether this is a NPO or a PO like in the other 
pitchfork bifurcations seen in \fig{bifs}), and $\sigma_0$ by
sign$(r-r_{b})$ using the appropriate bifurcation radius $r_{b}$. 
The Morse index $\mu_0$ in \eq{rho3unif} has to be chosen according 
the rules given in \cite{ss97}.

We must emphasize that the uniform approximation \eq{rho3unif}
only holds up to distances from the bifurcation point $r$, until
which the respective orbits do not change their nature by a new
bifurcation (or by symmetry restoring when going towards $r=0$).
This becomes a practical problem when successive bifurcations 
lie close, i.e., when the relevant action difference between the 
two points is of the order of, or smaller than $\hbar$. We discuss 
this problem again in the following section and in \sec{secnum}.

\subsubsection{Tangent bifurcations.}
\label{sectang}

The tangent bifurcations of POs in systems with $D=2$ always occur 
as generic isochronous bifurcations \cite{meye,then}. The two
orbits created thereby have infinite slopes of their stability
traces \cite{ss97,jaen1}:
\be
\lim_{\epsilon\to 0}\; {\rm Tr\,M}'(\epsilon) = \pm \infty\,.
\ee
The same holds for the Jacobians of the NPOs created by tangent 
bifurcations in our present system, as seen for the example of 
the orbits P and P' on the right side of \fig{bifs}. 
Characteristically, too, the Morse indices of the two orbits
differ by one unit close to the bifurcation (i.e., before any
further bifurcation occurs). 

The global uniform approximation for the tangent bifurcation has been
derived in \cite{ss97}; it looks similar to \eq{rho3unif} except
that the Bessel functions have the orders $\pm 1/3$ and $\pm 4/3$
(and can also be expressed in terms of Airy functions, cf.\
also \cite{bt}; on that side of the bifurcation where the orbits
are not real, modified Bessel functions occur). 

The reason for not quoting the result of \cite{ss97} here is
that for all NPOs in the circular billiard created in this way, 
the tangent bifurcation occurring at the critical point 
$r=r_{v,w}$ is immediately followed by a pitchfork bifurcation 
of one of the orbits at a nearby critical point 
$r=r_{v,w}^{\rm PO}>r=r_{v,w}$ where the PO $(v,w)$ is created 
(see the systematics in \sec{secnlin}). The two bifurcations 
lie close in the sense that the action difference 
$S_{v,w}(r=r_{v,w}^{\rm PO})-S_{v,w}(r=r_{v,w})$ is {\it not} 
much larger than $\hbar$. 
Consequently, the arguments of the Bessel functions in the
uniform approximations are never large enough to allow for
their asymptotic expansion. In other words: in the region
between the two bifurcations, the two orbits are never 
sufficiently isolated for the two bifurcations to be treated
separately. This necessitates to treat what in the relevant 
literature is called a ``codimension-two'' bifurcation. Such
bifurcations have been investigated and semiclassically
applied in \cite{mawu,schom}. We have, however, not been able 
to find an appropriate uniform approximation for our present 
bifurcation scenario which is similar to, but not identical 
with the ``butterfly catastrophe'' discussed in \cite{mawu}.
Luckily, the NPOs created by tangent bifurcations are  
practically negligible (cf.\ end of \sec{secconv}).

\newpage

\subsection{Regularization at the boundary $r=R$}
\label{secfried}

We have already mentioned in \sec{secNPOs} that the semiclassical
amplitude of the orbit L$_+^{(0)}$, which is responsible for the 
Friedel oscillations \cite{rbkm}, diverges at the boundary $r=R$ 
due to the zero of its Jacobian \eq{Jpm}. A method to regularize 
this divergence for the density using a uniform approximation of 
the Green function for short times was proposed in \cite{agam}. 
The explicit result for $D$-dimensional spherical billiards has 
been given in \cite{rbkm}; the particular result for $D=3$ was 
also obtained in \cite{bonc}. 
For the two-dimensional circular billiard we obtain 
the following uniform contribution of the primitive ``+'' orbit 
L$_+^{(0)}$ to the particle density:\footnote{Unfortunately, no 
corresponding uniform results for the kinetic energy densities
have been found. [Note also that the relations \eq{lvt} and 
\eq{tautau1} do not hold at and near the boundary $r=R$.]}
\begin{equation}
\delta\rho_+^{\rm (un)}(r) = -\frac{p_{\lambda}}{2\pi\hbar} 
                 \sqrt{\frac{R}{r}}\frac{J_1[2(R-r)p_{\lambda}/\hbar]}{(R-r)}\,.
\label{friedreg2d}
\end{equation}
Sufficiently far from the boundary, this result goes over into the correct
expression for the isolated L$_+^{(0)}$ orbit using \eq{amp}
and its properties given in \eq{Spm}, \eq{Jpm}:
\be
\delta\rho_+^{\rm (un)}(r) \longrightarrow
        -\frac{1}{2\pi(R-r)}\,\sqrt{\frac{p_\lambda R}{\pi\hbar r(R-r)}}\,
        \cos\left[\frac{1}{\hbar}\,S_+^{(0)}(r)-\frac34\,\pi\right].
\label{drplusiso}
\ee
\begin{figure}[t]
\begin{center}
\begin{minipage}{1.\linewidth}
\hspace{3cm}
\includegraphics[width=0.65\columnwidth,clip=true]{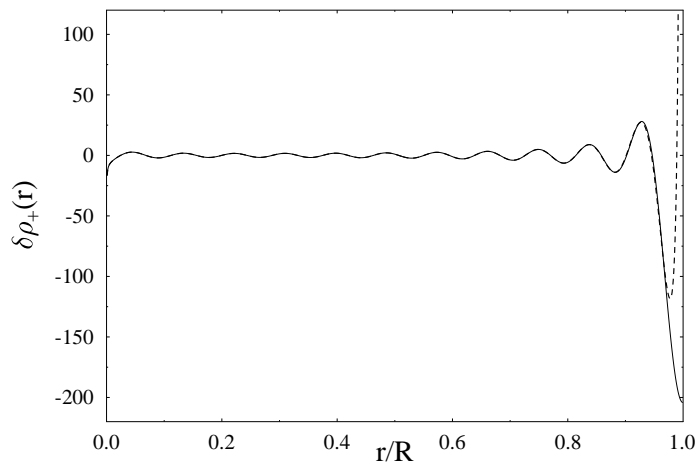}
\end{minipage}
\end{center}
\caption{
Contribution of primitive L$_+^{(0)}$ orbit to the particle density
in the circular billiard with $N=606$ particles ($\hbar^2/2m=R=1$).
{\it Solid line:} uniform approximation \eq{friedreg2d}, {\it 
dashed line:} asymptotic expression \eq{drplusiso}.
\label{fried}}
\end{figure}
This is illustrated in \fig{fried}. The ``recovery distance'' $d$ 
from the boundary, at which the diverging asymptotic result goes 
over into the uniform result, is bounded from below by the Fermi 
wave length:
\be
d \simg \hbar/p_\lambda\,.
\ee
For the particle number $N=606$ with $p_\lambda=35.8186$ chosen 
here, this corresponds to $d\simg 0.03$.

\newpage

\section{Numerical results for density oscillations}
\label{secnum}

In this section we present numerical calculations for the
spatial densities in the circle billiard and compare our
semiclassical results with exact quantum-mechanical results.
In \sec{secconv} we will discuss the convergence of the
sum over closed orbits in semiclassical expressions
\eq{drhosc} -- \eq{dtau1sc}, and in \sec{secshell} we address 
the role of closed main shells and investigate the relation 
of the spatial density oscillations with the shell effects 
in the total energy of the system.

\subsection{Convergence of orbit sums}
\label{secconv}

\begin{figure}[t]
\begin{center}
\begin{minipage}{1.\linewidth}
\hspace{3.3cm}
\includegraphics[width=0.75\columnwidth,clip=true]{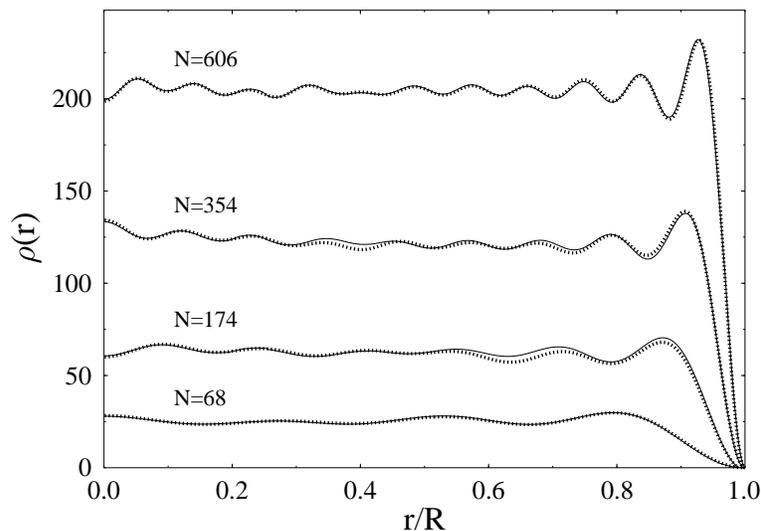}
\end{minipage}
\end{center}
\caption{\label{disk606}
Particle density in the two-dimensional disk billiard with $N=606$, 
354, 174 and 68 particles (units: $\hbar^2\!/2m=R=1$). The dotted lines are
the quantum results, the solid lines the converged semiclassical results 
using the number of orbits and the regularizations discussed in the text.
}
\end{figure}

Figure \ref{disk606} shows the total particle density $\rho(r)$ for 
four values of the number of $N$ particles  bound in the circular 
billiard. The dotted line is the quantum result \eq{rho}, and the 
solid line the converged semiclassical result obtained as explained
below. The numbers $N=606$ and 68 correspond to filled main shells 
(``closed-shell systems'') and the numbers $N=354$ and 174 to 
half-filled main shells (``mid-shell systems''); we refer to the 
discussion of the shell structure in \sec{secshell} (cf.\ \fig{decirc}).
The agreement for the closed-shell systems is excellent, that for 
the mid-shell systems is also very satisfactory.

\begin{figure}[t]
\begin{center}
\begin{minipage}{1.\linewidth}
\hspace{3.3cm}
\includegraphics[width=0.64\linewidth,clip=true]{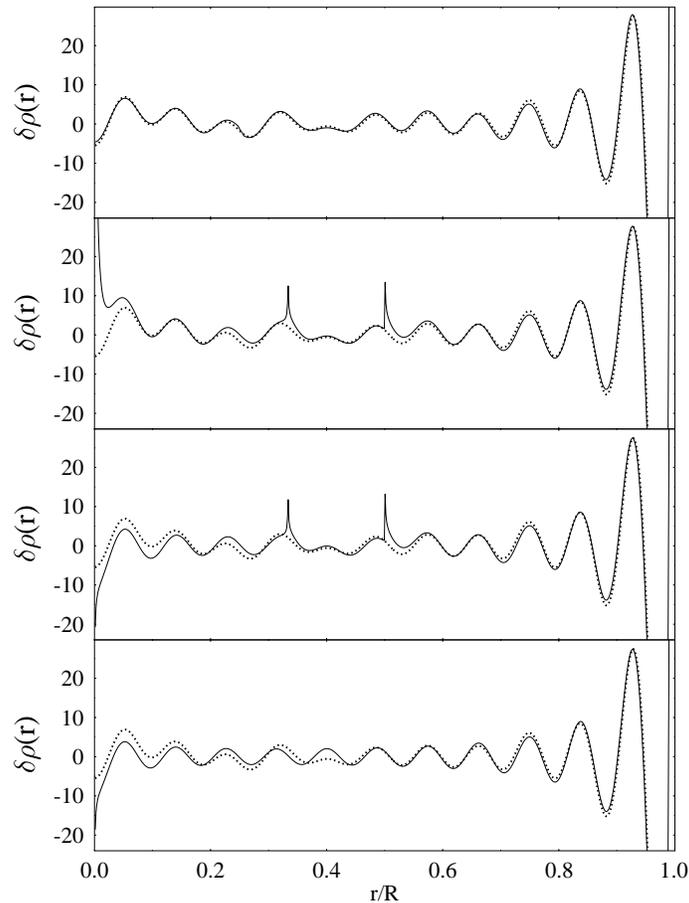} %\vspace*{-0.2cm}
\end{minipage}
\end{center}\vspace*{-0.5cm}
\caption{\label{figdisk303}
Oscillating part $\delta\rho(r)$ of the particle density for 
$N=606$ particles like in \fig{disk606}. {\it Dots:}
quantum results (identical in all panels). {\it Solid lines:}
semiclassical results; in the three lowest panels as isolated orbits 
with amplitudes \eq{amp}, and in the top panel using the uniform 
approximations \eq{rhorunif}, \eq{rho2kkunif} and \eq{rho3unif}
(see text for details).
}
\end{figure}

In \fig{figdisk303} we show how the convergence
of the semiclassical result for $N=606$ comes about; here only 
the oscillating part $\delta\rho(r)$ of the density is shown.
Dotted lines show the quantum results, identical in all panels, 
and solid lines the semiclassical results in various
approximations. From bottom to top, an increasing number of 
closed orbits is included in the sum \eq{drhosc}. In the three 
lowest panels, the amplitudes ${\cal A}_\gamma$ \eq{amp} for 
isolated orbits are used; in the top panel, the uniform 
approximations for the symmetry breaking at $r=0$ and for 
the pitchfork bifurcations are used.

In the bottom panel, 
only the primitive radial NPOs\, L$_\pm^{(k=0)}$ 
are included. In spite of their weak divergence at $r=0$, their
contribution already accounts for most of the density oscillations.
In particular, the Friedel oscillations are well reproduced by the
``+'' orbit up to the highest peak; only close to the boundary its
contribution diverges (cf.\ the contribution of the ``+'' orbit
alone in \fig{fried}). In the second panel from below, the radial
orbits L$_\pm^{(k=1)}$ have been added, together with the non-radial 
orbit $\Lambda$ (3,1) and the triangle PO (3,1) emerging from 
L$_+^{(k=1)}$ by pitchfork bifurcations at $r=R/3$ and $r=R/2$,
as shown in \fig{bifs} (left panels). The agreement with the quantum
results is slightly improved in the region $0.3 \siml r/R \siml 0.6$,
but at the cost of the divergences at the two bifurcation points.
In the third panel, we have furthermore added the non-radial orbit 
$\nabla$ (2,1) and the diameter PO (2,1). This improves the agreement
around $0.1 \siml r/R \siml 0.2$, but at the cost of a stronger
divergence at $r=0$. In the top panel, finally, we have used the
uniform approximation \eq{rhorunif} for the symmetry breaking
at $r=0$ of the ``+'' and ``$-$'' orbits, including also the radial 
orbits L$_\pm^{(k=2)}$ and the orbits $(2k+1,k)$ bifurcating 
from them, the uniform approximation \eq{rho3unif} for the 
pitchfork bifurcations of the NPOs and POs (3,1) and (5,2), and
finally the uniform approximation \eq{rho2kkunif} for the symmetry 
breaking at $r=0$ of the NPOs and POs $(2k,k)$ including values of 
$k$ up to $k=10$. The agreement with quantum mechanics is now
excellent, except very close to the boundary where the isolated
``+'' orbit (with $k=0$) diverges. Employing the regularization
at $r=R$ discussed in \sec{secfried} and shown separately in
\fig{fried} leads to the result shown in \fig{disk606}.
\begin{figure}[t]
\begin{center}
\begin{minipage}{1.\linewidth}
\hspace{2cm}
\includegraphics[width=.72\linewidth,clip=true]{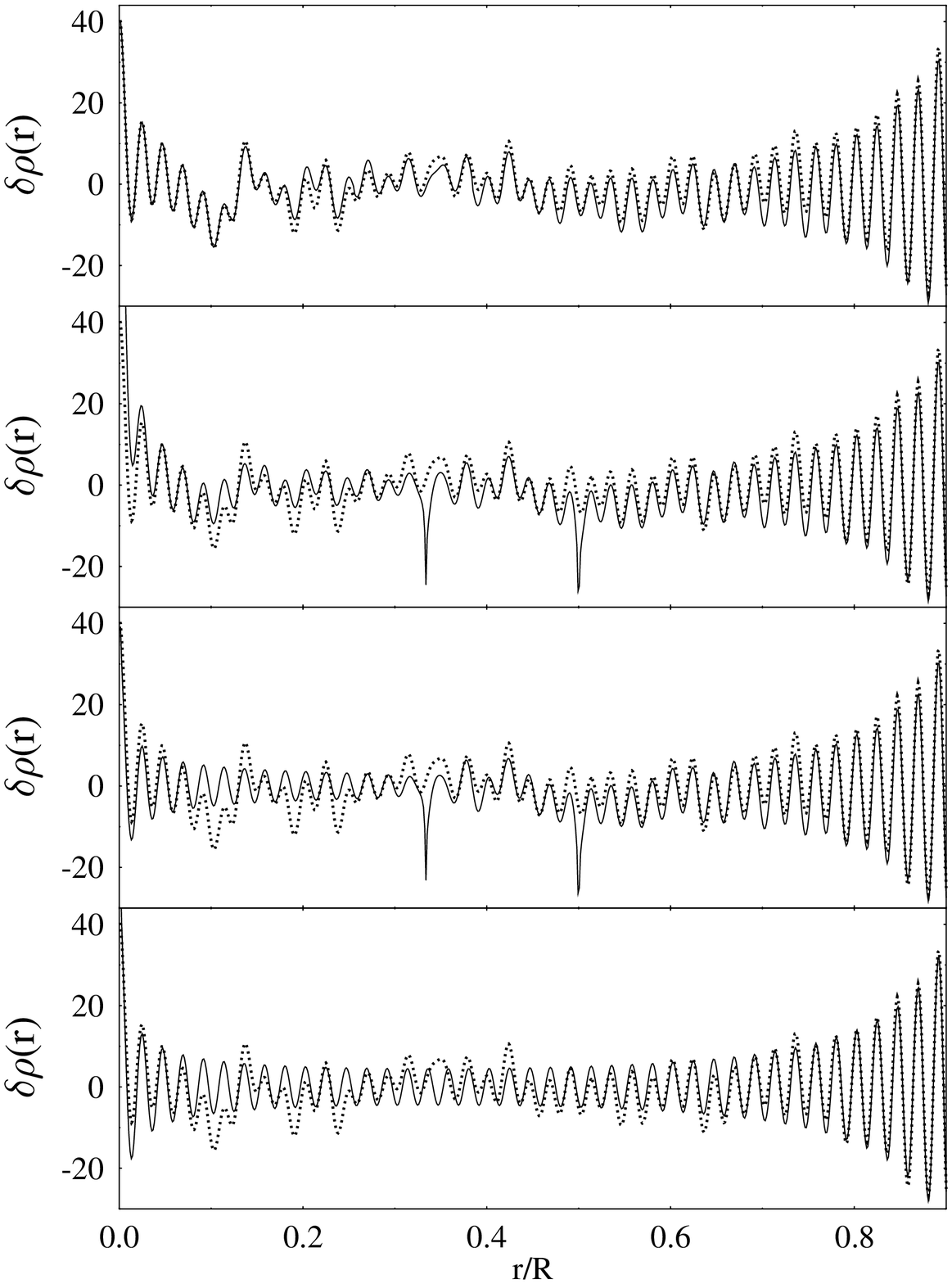}
\end{minipage}
\end{center}\vspace*{-0.3cm}
\caption{\label{figdisk4917}
Same as \fig{figdisk303}, but for $N=9834$ particles.
}
\end{figure}

The same analysis is made in \fig{figdisk4917} for a larger
system with $N=9834$ particles; the convergence is similar
as above. Here the irregular part of the oscillations is seen 
to be well reproduced by the contributions of the non-radial 
NPOs $(2k,k)$ in the two upper panels. We have omitted the
surface region $r\geq 0.9R$ because the peaks of the last 
Friedel oscillations are so high that the oscillations 
in the interior would not be recognizable on their scale.
The contribution of the primitive ``+'' orbit alone, however,
reproduces the quantum density almost exactly for $r>0.9R$ 
like in \fig{disk606}.
\begin{figure}[t]
\begin{center}
\begin{minipage}{1.\linewidth}
\hspace{3.3cm}
\includegraphics[width=.72\linewidth,clip=true]{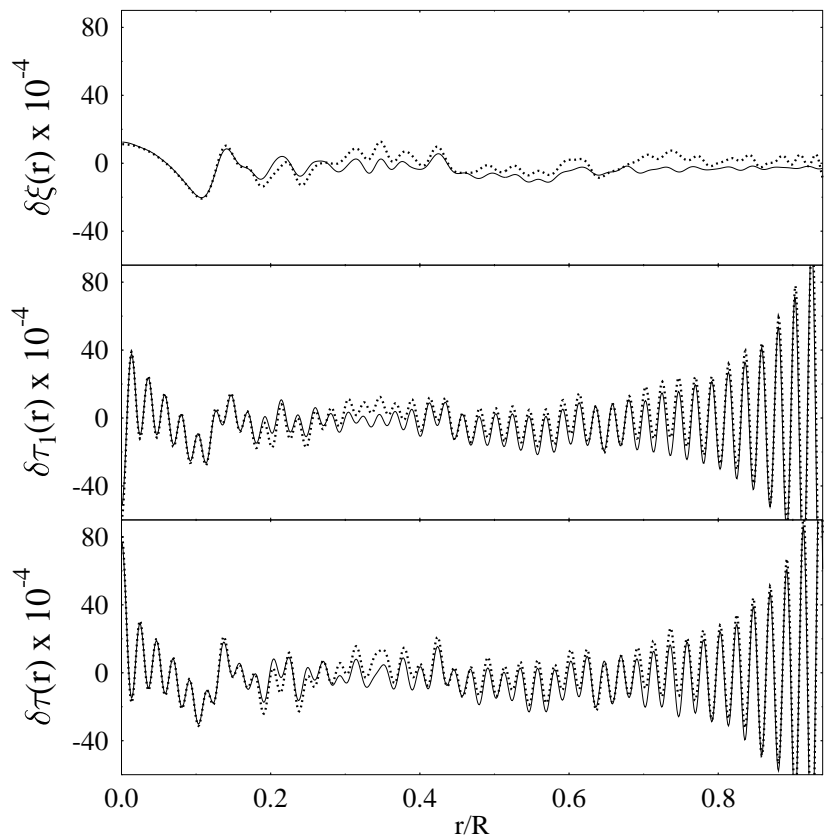}
\end{minipage}
\end{center}\vspace*{-0.5cm}
\caption{\label{2ddisk4917}
Quantum-mechanical results (dots) and uniform semiclassical results (lines)
for $N=9834$ particles as in top panel of \fig{figdisk4917}, but for
the oscillating parts of the kinetic energy densities $\delta\xi(r)$ (top panel), 
$\delta\tau_1(r)$ (center panel), and $\delta\tau(r)$ (bottom panel).
}
\end{figure}

In \fig{2ddisk4917} we show the oscillating parts of the kinetic
energy densities of the system with $N=9834$ particles using the
same orbits and uniform approximations as in the top panel of 
\fig{figdisk4917}. Like there, the agreement is best in the region 
$0\leq r \siml 0.2R$, where the rapid oscillations are dominated by 
the radial orbits L$_\pm^{(k)}$ and the slow, irregular oscillations 
by the orbits $(2k,k)$. We see that the rapid oscillations are opposite 
in phase in the two kinetic energy densities $\delta\tau(r)$ and 
$\delta\tau_1(r)$, as expressed in the relation \eq{tautau1} which is 
due to the fact that the momentum mismatch for all radial orbits has 
the constant value $Q=-1$. The rapid oscillations cancel therefore in 
the quantity $\delta\xi(r)$, seen in the top panel of \fig{2ddisk4917}, 
which is due exclusively to the non-radial orbits (except very close to 
the boundary) as discussed in general in \cite{rb}. 

The small discrepancies at $r\simg 0.2R$ are due to missing longer 
orbits in the sums contributing to \eq{drhosc} -- \eq{dtau1sc}. 
In principle, these can be included using the general formulae given 
in \sec{secscl}. There is a practical problem, however. With
increasing values of $k$, the bifurcations tend to lie denser 
along the radial variable $r$. The uniform approximations given
in \sec{secregul}, including those for the symmetry breaking at
$r=0$, can only be used for the regularization of a divergence around 
one critical point at a time. It is necessary that, while going away 
from one critical point, an asymptotic domain be reached in which 
the uniform approximation becomes identical to the sum of
contributions of the isolated orbits, before one approaches the
next critical point. Midways between the two critical points, one
can then switch from one uniform approximation to the next one.
This is what we have done in the examples given in the figures of 
this section; it was possible with the limitation of $k_{\rm max}=2$ 
for the bifurcating orbits of type $(2k+1,k)$. For longer orbits,
successive bifurcations lie so close that their actions at the
two bifurcation points differ by less than $\sim \hbar$ and hence
the bifurcations cannot be treated separately. This is the same
problem as arises for the pairs of orbits created by tangent 
bifurcations (see e.g.\ \fig{bifs}, right panels), as discussed
in \sec{sectang}. We can consider it as a lucky circumstance that, 
with the quality of the semiclassical approximation reached in the 
above figures, all those longer orbits can be practically neglected. 

\subsection{Relation to shell effects in the total energy}
\label{secshell}

\begin{figure}[h]
\begin{center}
\begin{minipage}{1.\linewidth}
\hspace{2.5cm}
\includegraphics[width=.75\linewidth,clip=true]{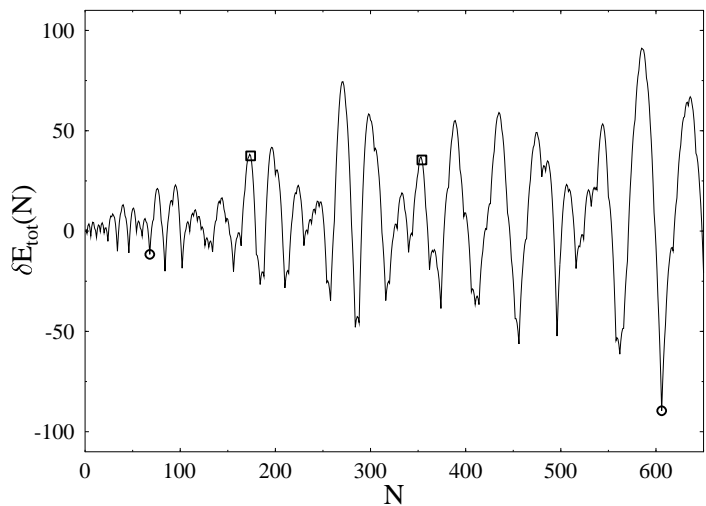} 
\end{minipage}
\end{center}\vspace*{-0.3cm}
\caption{\label{decirc}
Shell-correction energy $\delta E_{\rm tot}(N)$ \eq{delta-E} versus particle 
number $N$ for the circle billiard (energy units: $E_0=1$). The circles
mark the the closed-shell systems (at minima of $\delta E_{\rm tot}$) with 
$N=68$ and 606, and the squares mark the mid-shell systems (near maxima of 
$\delta E_{\rm tot}$) with $N=174$ and 354 tested in \fig{disk606}.
}
\end{figure}

As already observed in \fig{disk606}, the quality of the
semiclassical approximation depends to some extent on the shell 
situation of the system with $N$ particles. This is known also from 
semiclassical calculations of the total energy of finite fermionic 
systems \cite{strm,mbc,lm}. To study the shell structure, it is 
instructive to consider the {\it shell-correction energy} 
$\delta E_{\rm tot}$ defined \cite{stru} as the oscillating part of the 
total energy: 
\bea
\hspace{-2.cm}
\delta E_{\rm tot} = 2\sum_{E_j\leq\lambda}\!\! E_j\,
                 -\,2\!\int_0^{\lambdab}\!\! E\,\ggt(E)\,\d E% \nonumber\\
         = 2\!\int_0^{\lambda}\!\!E\,g(E)\,\d E -2\!\int_0^{\lambdab}\!\!E\,\ggt(E)\, \d E\,. 
\label{delta-E}
\eea
Here $g(E)=\sum_j\delta(E-E_j)$ is the exact level density and $\ggt(E)$ 
its average part. $\lambda$ and $\lambdab$ are the Fermi energies 
determined by $N={\cal N}(\lambda)={\widetilde{\cal N}}(\lambdab)$ in 
terms of the ``number counting functions'', which are defined as the first 
integrals of the two level densities:
\bea
{\cal N}(E) & = & 2\sum_j\Theta(E-E_j) = 2\!\int_0^E \!\!g(E)\,\d E\,,\nonumber\\
{\widetilde{\cal N}}(E) & = & 2\!\int_0^E\!\! \ggt(E)\, \d E
     = \frac{E}{2E_0}-\sqrt{\!\frac{E}{E_0}}+\frac13+{\cal O}(E^{-1/2})\,.
\label{NofE}
\eea
The explicit expression for ${\widetilde{\cal N}}(E)$ is obtained 
from the Weyl expansion \cite{bh,beho}.
In \fig{decirc} we show the shell-correction energy as a function of
the particle number $N$ in the region $0\leq N \leq 650$. A very 
pronounced shell structure is seen. The minima of $\delta E_{\rm tot}(N)$ 
correspond to closed ``main shells'' for which the level density at the 
Fermi energy has a minimum, too, corresponding to a low degeneracy of 
the spectrum. These main shells may be counted by the number $M$
appearing in \eq{delrhorad}. The maxima of $\delta E_{\rm tot}(N)$ 
correspond to regions of high degeneracy for ``mid-shell'' systems. 
This behavior of the energy shell structure has been studied 
extensively, e.g., in nuclei \cite{fuhi}.

In order to calculate the shell-correction energy $\delta E_{\rm tot}$
semiclassically, it is convenient to express it directly in terms
of the oscillating part $\delta g(E) = g(E)-\ggt(E)$ of the level
density, for which a Gutzwiller-type trace formula \cite{gutz} can be 
employed. $\delta E_{\rm tot}(N)$ may, in fact, be approximated \cite{strm,lm}
by either of the two following expressions
\begin{equation}
\delta E_{\rm tot}(N)\, \simeq \int_0^\lambda (E-\lambda)\,\delta g(E)\,\d E\,
                    \simeq \, -\!\int_0^{\lambdab}\delta{\cal N}(E)\,\d E\,,
\label{delta-E-app}
\end{equation}
where $\delta{\cal N}(E)={\cal N}(E)-{\widetilde{\cal N}}(E)$ is the
oscillating part of the number counting function. The two relations
given in \eq{delta-E-app} are correct up to terms of order $(\lambda
-\lambdab)^2$ or $[\delta{\cal N}(\lambdab)]^2$, respectively. The 
semiclassical trace formula for $\delta g(E)$ of the circular 
billiard is given in \cite{disk}. After its integration in 
\eq{delta-E-app}, one obtains to leading order in $\hbar$ the
following semiclassical approximation for the shell-correction energy
in terms of the POs $(v,w)$
\be
\hspace{-2.5cm}
\delta E_{\rm tot}(N)\, \simeq  2\, E_0^{1/4}\, \lambdab^{3/4} \mathop 
                            \sum_{w=1}^\infty\sum_{v=2w}^\infty 
                            \frac{f_{v,w}}{v^2\sqrt{\pi v\sin(w\pi/v)}}\,
                            \sin\!\left[\frac{1}{\hbar}p_\lambda L_{v,w}^{\rm PO}
                            -3v\frac{\pi}{2}+\frac{3\pi}{4}\right]\!,
\label{delta-E-sc}
\ee
where the lengths $L_{v,w}^{\rm PO}$ are given in \eq{lpo} and 
$f_{v,w}=2-\delta_{v,2 w}$.
\begin{figure}[h]
\begin{center}
\begin{minipage}{1.\linewidth}
\hspace{3cm}
\includegraphics[width=0.8\linewidth,clip=true]{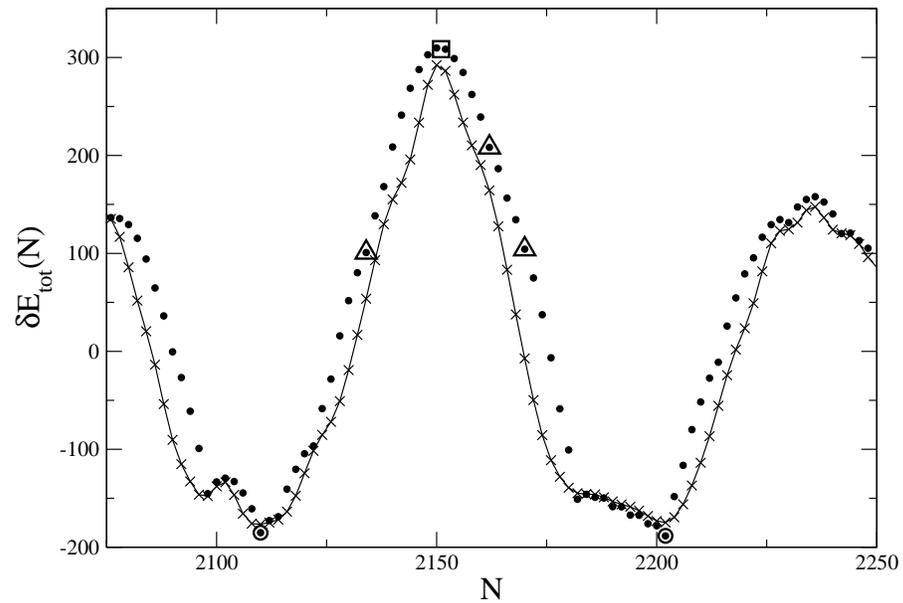} 
\end{minipage}
\end{center}\vspace*{-0.3cm}
\caption{\label{dE-N}
The same as \fig{decirc} in a different region of particle numbers $N$.
{\it Dots:} exact values \eq{delta-E}, {\it crosses connected by the 
solid line:} semiclassical values \eq{delta-E-sc}. The circles mark 
closed-shell systems, the square marks mid-shell systems, and the
triangles mark some intermediate systems tested in \fig{badcirc} below.
}
\end{figure}

In \fig{dE-N} we compare the semiclassical results \eq{delta-E-sc} of 
$\delta E_{\rm tot}(N)$ with their exact values \eq{delta-E} in the region 
$2075\leq N \leq 2250$. We see that the agreement is best at the minima 
(closed main shells) and fairly good around the maxima (mid-shell systems). 
In regions where $\delta E_{\rm tot}(N)$ has steep slopes, there are
relatively large discrepancies, as already observed earlier \cite{mbc,lm}. 
The reason lies in the fact that in these regions, the stair-case functions 
$\lambda(N)$ and ${\cal N}(E)$ have their largest deviations from the 
smooth functions $\lambdab(N)$ and ${\widetilde{\cal N}}(E)$, so that
the missing terms of order $(\lambda-\lambdab)^2$ or $[\delta{\cal 
N}(\lambdab)]^2$ are largest, while these are smallest at closed main 
shells and near mid-shell regions.
\begin{figure}[t]
%\vspace*{0.2cm}
\begin{center}
\begin{minipage}{0.95\linewidth}
\hspace{2.5cm}
\includegraphics[width=0.8\linewidth,clip=true]{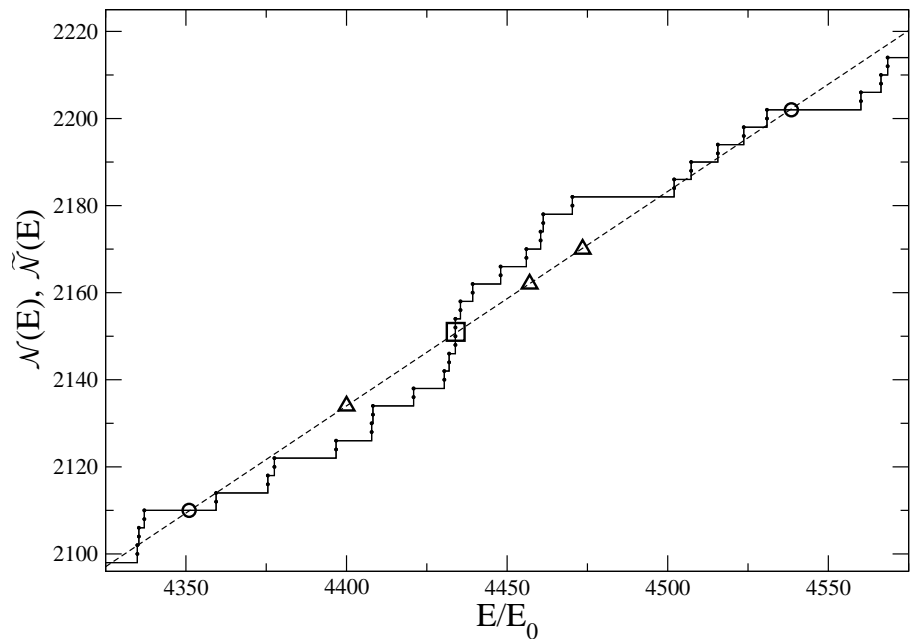} 
\end{minipage}
\end{center}\vspace*{-0.3cm}
\caption{\label{lamofN}
Number counting functions ${\cal N}(E)$ ({\it dots connected by the 
solid line}) and ${\widetilde{\cal N}}(E)$ (dashed line).
The circles, triangles and the square correspond to those in \fig{dE-N}
in the region $2098 \le N \le 2222$.
}
\end{figure}
This is illustrated in \fig{lamofN}, where we show the two
curves ${\cal N}(E)$ and ${\widetilde{\cal N}}(E)$ versus 
energy $E$ in an interval covering the central part of the 
region of particle numbers shown in \fig{dE-N}. They are seen to 
intersect at closed shells (circles) and in the mid-shell region 
(square). The triangles mark ``difficult systems'' for which the 
errors in the semiclassical results are largest, as seen in 
\fig{dE-N}. There is a clear correlation between these errors
and the magnitude of $\delta{\cal N}(E)$.
\begin{figure}[h]
\begin{center}
\begin{minipage}{1.\linewidth}
\hspace{3cm}
\includegraphics[width=.75\linewidth,clip=true]{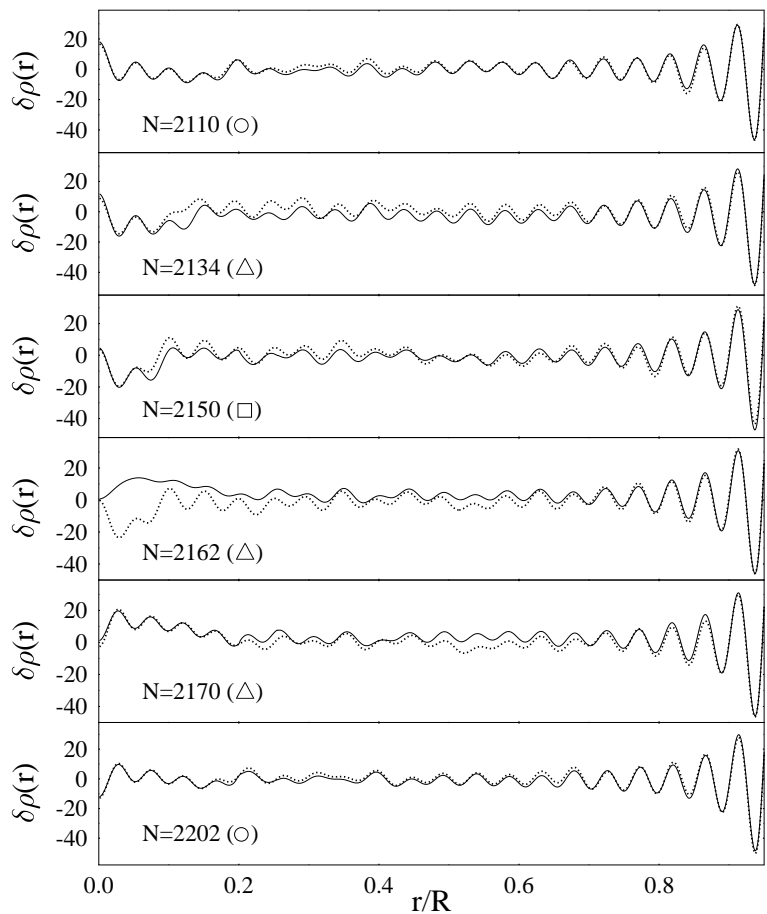} 
\end{minipage}
\end{center}\vspace*{-0.5cm}
\caption{\label{badcirc}
Oscillating part $\delta\rho(r)$ of the spatial density, 
obtained like in the top panels of figures \ref{figdisk303} and 
\ref{figdisk4917}, here for the six particle numbers marked by 
the circles, triangles and the square in \fig{lamofN}.
}
\end{figure}

We observe a similar trend for the quality of our semiclassical 
approximation to the spatial density oscillations. The four 
systems shown in \fig{disk606}, where the agreement with the 
quantum results is best, correspond to closed-shell and mid-shell 
systems (see the triangles in \fig{decirc}). In \fig{badcirc} 
we show the corresponding results for the five systems marked by
circles, triangles and the square in \fig{lamofN}. Clearly, the
best agreement is reached for the closed-shell systems (circles).
The correlation of the errors in $\delta\rho(r)$ with the 
magnitudes of $\delta{\cal N}(E)$ is less direct than for the
shell-correction energy, but the agreement in \fig{badcirc} is 
clearly worst in the systems (marked by triangles) for which 
$\delta E_{\rm tot}(N)$ has its steepest slope. The mid-shell
system $N=2150$ here has slightly larger errors (like also
for $\delta E_{\rm tot}$ in \fig{dE-N}) than those shown in the
middle panels of \fig{disk606}. The same trends are also found
for the kinetic energy densities. 

In conclusion we can state that, with a small number of included
classical orbits and straight-forward uniform approximations, the
semiclassical approximation for the spatial density oscillations
in the circular billiard is excellent for closed-shell systems, 
good for mid-shell systems and still quite satisfactory for most 
intermediate systems.

\section{Summary and outlook}
\label{secsum}

We have presented a complete classification of the closed periodic 
and non-periodic classical orbits in the two-dimensional circular 
billiard with radius $R$. We have divided the orbits into two classes: 
radial orbits, librating in the radial directions, and non-radial 
orbits. Analytical formulae are given for their actions, stability 
determinants, momentum mismatches and Morse indices. We have found 
that many of the orbits can undergo bifurcations at which new orbits 
are created. The bifurcation parameter hereby is their starting point 
$r$ varied in the range $0<r<R$. Periodic orbits can be created from 
non-periodic ones, and {\it vice versa}, in isochronous pitchfork 
bifurcations. Pairs of non-periodic orbits can also be generated from 
tangent bifurcations. Finally, a class of non-periodic orbits is 
generated from the radial periodic orbit upon breaking its rotational 
U(1) symmetry by varying its starting point from $r=0$ to $r>0$.

Employing a recently developed closed-orbit theory \cite{rb,rbkm}, 
we have investigated the semiclassical approximation of spatial 
density oscillations for $N$ non-interacting fermions bound in 
the circular billiard in terms of its closed orbits. At the 
critical points (symmetry breaking at $r=0$, bifurcation points,
and the boundary $r=R$), uniform approximations have to be used
in order to regularize their semiclassical amplitudes. For the
symmetry breaking and both two types of bifurcations,
standard uniform approximations from perturbation and bifurcation
theory could be implemented in a straightforward manner, having
asserted that the bifurcations are exactly of the types known
for periodic orbits as functions of other bifurcation parameters
(such as energy, deformation or external field strength).
Following the general analysis in \cite{rb,rbkm}, we have
demonstrated that also in the present system the radial orbits 
are responsible for the short-ranged regular oscillations in both 
particle and kinetic-energy densities, while the non-radial orbits 
create their long-ranged irregular oscillations. We have also 
confirmed here that the radial orbit with one reflection at the 
nearest turning point creates the Friedel oscillations near the 
boundary.

We have finally tested the semiclassical approximation by comparing 
its results to exact quantum-mechanical densities. We investigated 
the convergence of the sum over closed orbits and the correlations 
of its results with the shell structure in the total energy. We find 
that, using a relatively small number of included classical orbits 
and easy analytical uniform approximations, the semiclassical 
approximation is excellent for closed-shell systems, good for 
mid-shell systems and still quite satisfactory for most intermediate 
systems. This is found to hold even for moderate particle numbers 
$N\simg 50$, in spite of the fact that {\it a priori} the
semiclassical limit of a small Planck constant $\hbar$ corresponds 
to the limit of large particle numbers.

The classification of closed orbits that we have given here is
valid for spherical billiards in arbitrary
dimensions $D\geq 2$, since the motion always takes place in a
two-dimensional plane due to the conservation of angular momentum.
The only difference from our present results for $D>2$ is the
degeneracy of the orbits: all non-radial NPOs starting at $r>0$
have at least one continuous symmetry of rotation and are therefore 
no longer isolated; also the degeneracies of the POs will change
\cite{strm,crli}. 
While the shapes and actions of all orbits remain the same in all 
dimensions, their Jacobians and Morse indices become different for 
$D>2$, and the regularisation schemes discussed in \sec{secregul} 
have to be modified. This work is actually in progress. 

For smooth potentials with radial symmetry, we can expect all
types of orbits found in the present system, except that the
non-radial orbits will have smoothed reflections from the
potential wall instead of corners. But there is reason to
expect further orbits. There always exists a circular orbit 
(not present in a billiard) whose radius is given by the
minimum of the effective potential including the centrifugal
part. Further orbits are rather likely to be created by 
bifurcations, although we must expect the bifurcation scenarios
to be different from those found here. In \cite{arit} it has, e.g., 
been shown that in radial potentials of the form $V(r)=c\,r^\alpha$,
bifurcations of the circular orbit give birth to the (smoothed)
polygon-like POs under variation of $\alpha$.

For arbitrary potentials without symmetries, it is in general
rather hard to predict the shapes and properties of the classical 
orbits. In \cite{rbkm} we have discussed the applicability of our 
semiclassical closed-orbit theory to arbitrary potentials to some
extent, and also given explicit results for the rectangular billiard 
in which all closed orbits are easily classified. For potentials
with arbitrary shapes (smooth or hard-reflecting), one will in 
general have to rely on the numerical search of closed orbits and 
their bifurcations.

\bigskip

\ack
\addcontentsline{toc}{section}{\protect\numberline{ }Acknowledgements}
We acknowledge stimulating discussions with A G Magner and M Guti\'errez.

\newpage

\appendix
\setcounter{section}{1}
\section*{Appendix. Explicit results for specific non-radial NPOs}
\label{secapp}
%\markboth{{\it  Appendix}}{{\it References }}
\addcontentsline{toc}{section}{\protect\numberline{ }Appendix}
All the properties of the NPOs used in the semiclassical expressions 
\eq{drhosc} -- \eq{dtau1sc} for the density oscillations have been given
in the general analytical expressions in \sec{secNPOs} in terms of the 
angles $\alpha$ and $\beta$ related by \eq{betanpo} and shown in 
\fig{circle}. In practice, however, we need them as functions of the 
radial variable $r$. Unfortunately, for the non-radial NPOs the
inversion of the relation \eq{rofal} is not possible in general; using 
the standard relations of the trigonometric functions this would
require to find the roots of Chebychev polynomials. In the following, 
we give explicit results as functions of $r$ for some simple orbits, 
for which the explicit inversion of \eq{rofal} is possible. Some of 
these results have been used in examples given in the main part of the 
paper. 

We recall that the properties of the radial NPOs are given explicitly 
in \sec{seclin}; the general formulae for the Morse indices of the 
non-radial NPOs are given in the systematics of \sec{secnlin}.

\subsection{$v=2$, $w=1$: the triangle orbit $\nabla$ {\rm (2,1)}}
\label{secdel2}

For this orbit, which exists at all starting points $0<r\leq R$
(see \fig{nabla2} for its shape), the relation \eq{rofal} becomes
\be
r(\alpha)=R\,\frac{\sin\alpha}{\cos{(2\alpha})} = R\,\frac{x}{(1-2x^2)}\,,
\qquad x = \sin\alpha\,.
\ee
Inverting the above we get
\be\hspace{-1.cm}
x(r) = \frac{\sqrt{R^2+8r^2}-R}{4r}\,, \qquad
\alpha(r) = {\rm arcsin}\left(\frac{\sqrt{R^2+8r^2}-R}{4r}\right).
\label{xalofr}
\ee
The length of the $\nabla$ orbit becomes
\be
L_{2,1}(r) = 4R\,\frac{\cos^3\alpha}{\cos(2\alpha)} 
           = 4R\,\frac{\,(1-x^2)^{3/2}}{(1-2x^2)}\,,
\label{L2}
\ee
and the Jacobian becomes
\be
{\cal J}_{2,1}(r) = \frac{R}{p_\lambda}\tan(2\alpha)\!
                    \left[\sin\alpha+\frac{r}{R}\,(1+6\sin^2\!\alpha)\right].
\label{J2}
\ee
Inserting $\alpha(r)$ from \eq{xalofr} into \eq{L2} and \eq{J2}, both
quantities are obtained explicitly as functions of $r$. For small $r$
we get the Taylor expansions
\be\hspace{-1.7cm}
L_{2,1}(r) = 4R + 2\,\frac{r^2}{R} +  {\cal O}\left(\frac{r}{R}\right)^4,\qquad 
{\cal J}_{2,1}(r) = 4\,\frac{R}{p_\lambda}\left(\frac{r}{R}\right)^2
                    + {\cal O}\left(\frac{r}{R}\right)^4.
\label{expLJ2}
\ee

\subsection{$v=3$, $w=1$: the orbit $\Lambda$ {\rm (3,1)}}
\label{seclam3}

This orbit exists only for $r\geq r_1=R/3$, where it bifurcates 
from the radial orbit L$_+^{(1)}$. At $r=r_{3,1}^{\rm PO}=R/2$ the 
triangular PO $\Delta$ (3,1) bifurcates from it (see the bifurcation
scenarios in the left panels of \fig{bifs} and the shapes of the orbit
in \fig{Lamshapes}). At the boundary $r=R$ it becomes 
identical with the squared PO $\square$ (4,1). The relation between 
$r$ and $\alpha$ is given by
\be
r(\alpha)=R\,\frac{\sin\alpha}{\sin{(3\alpha})} = R\,\frac{1}{(3-4x^2)}\,.
\label{rofal31}
\ee
Inverting the above, we get
\be
\sin\alpha(r) = x(r) = \frac12\,\sqrt{3-\frac{R}{r}}\,,  \qquad r \geq r_1=R/3\,.
\label{x31}
\ee
The explicit expressions for its length, momentum mismatch and
Jacobian are
\bea
L_{3,1}(r) & = & 2\sqrt{\frac{R+r}{r}}\,(R+r)\,,\label{L3}\\
Q_{3,1}(r) & = & \frac{R^2}{2r^3}\,(3r-R)-1\,,\label{Q3}\\
{\cal J}_{3,1}(r) & = & \frac{2}{Rp_\lambda}\,(2r-R)(3r-R)\sqrt{\frac{r+R}{r}}\,.
\label{J3}
\eea
Note that, although the relations \eq{rofal31} and \eq{x31} only 
hold for $r\geq r_1$, the three functions \eq{L3} -- \eq{J3}  can be 
continued analytically to $r<r_1$ and are real in the entire interval 
$0\leq r\leq R$. For $r<r_1$, we can consider them as the (real) 
properties of an imaginary ``ghost orbit''; they are required in the 
uniform approximation \eq{rho3unif} for the pitchfork bifurcation at 
$r_1$ discussed in \sec{secpitch}.

\subsection{$v=4$, $w=1$: the ``pentagon'' orbits {\rm P (4,1)} and {\rm P' (4,1)'}}
\label{secp41}

This pair of NPOs is created in a tangent bifurcation at
\be
r_{4,1}=0.682489R\,.
\label{r41}
\ee
Some of their shapes are shown in \fig{P4shapes}, and their
bifurcation scenario is shown in the right panels of \fig{bifs}.
At the boundary $r=R$, P becomes identical to the pentagonal
(not self-crossing) PO (5,1), and P' becomes identical to the
triangular PO $\Delta$ (3,1).
The connection between $r$ and $\alpha$ of this orbit is given by
\be
r(\alpha)=-R\,\frac{\sin\alpha}{\cos(4\alpha)} = -R\,\frac{x}{(1-8x^2+8x^4)}\,.
\label{rP4}
\ee
The inversion of $r(x)$ leads to a fourth-order algebraic equation 
whose analytic solution is possible but cumbersome. Without writing 
it down explicitly, we can make the following statements. The real 
solutions for $x(r)$ consist of two branches starting from the 
minimum value of $r(x)$ which is found at 
$x_0=[(2+\sqrt{10})/12]^{1/2}=0.655889$ and has the value 
$r(x_0)=r_{4,1}$ given in \eq{r41}. The two 
branches correspond to the two orbits P and P' born in a tangent 
bifurcation at $r_{4,1}$. The orbit P' corresponds to the lower 
branch with $x'_1=1/2\leq x \leq x_0$. The orbit P corresponds to 
the upper branch with $x_0 \leq x \leq x_1=\sin(3\pi/10)=0.809017$. 
At $r_{4,1}^{\rm PO}=R/\!\sqrt{2}$ it undergoes a pitchfork
bifurcation, whereby the squared PO $\square$ (4,1) is born. The lengths,
momentum mismatches and Jacobians can in principle be evaluated
explicitly but are most easily calculated numerically using the
general formulae given in \sec{secNPOs}. We note that the analytical
continuation to $r<r_{4,1}$, where the orbits are imaginary ``ghost 
orbits'', here leads to complex quantities, which is a characteristic 
of the tangent bifurcation \cite{ss97}.

For the sake of an illustration, we give in the following the results
for the similar pair of NPOs with $(v,w)=(5,1)$, for which the two 
branches of $x(r)$ can be given explicitly.

\subsection{$v=5$, $w=1$: the pair of orbits {\rm (5,1)} and {\rm (5,1)'}}

These orbits are born in a tangent bifurcation at the point $r_{5,2}=4R/5$.
Inverting
\be
r(\alpha) = -R\,\frac{\sin\alpha}{\sin(5\alpha)} = \frac{-R}{(5-20x^2+16x^4)}\,,
\ee
which has a minimum at $x_0=\sqrt{5/8}$, yields a bi-quadratic equation
with the simple solutions
\be
\sin\alpha(r) = x(r) = \frac12\,\sqrt{\frac12\left(5\pm\sqrt{5-4R/r}\right)}\,,
\ee
whereby the two signs correspond to the two orbits. Their properties are
\bea\hspace{-1.cm}
L_{5,1}(r) =  R\sqrt{\frac12\left(3\mp\sqrt{5-4R/r}\right)}
                 \left(4+\frac{r}{R}\pm\frac{r}{R}\sqrt{5-4R/r}\right),\label{L51}\\
\hspace{-1.cm}Q_{5,1}(r)  =  -1+\frac{R^2}{4r^2}\left(5\mp\sqrt{5-4R/r}\right),\label{Q51}\\
\hspace{-1.cm}{\cal J}_{5,1}(r)  =  \frac{1}{Rp_\lambda}\sqrt{\frac12\left(3\mp\sqrt{5-4R/r}\right)}\nonumber\\
\hspace{-1.cm}     \hspace{1.6cm}   \times  \left[4R^2+30r^2-29Rr\pm r(-9R+10r)\sqrt{5-4R/r}\right]\!.
\label{J51}
\eea
The upper and lower signs in the above expressions correspond to
the orbits (5,1) and (5,1)', respectively. At $r_{5,1}^{\rm
  PO}=R\,\cos(\pi/5)$, which is a zero of the term in square brackets
on the second line in \eq{J51} (taken with the upper sign), the NPO (5,1)
gives birth to the regular pentagon PO (5,1) in a pitchfork bifurcation.

\end{document}